\documentclass[aip,jmp,preprint]{revtex4-1}
\usepackage{amsmath}
\usepackage{amsfonts,amssymb}
\usepackage{bm}
\usepackage{color}

\draft % marks overfull lines with a black rule on the right

\renewcommand{\vec}[1]{ \boldsymbol{#1} }
\renewcommand{\Re}[1]{ \mathrm{Re}{#1} }

\def\G{{\mathbf {\Gamma}}}
\def\P{{\bf P}}
\def\vp{\vec{p}}
\def\A{{\mathcal A}}

\def\bPh{\bm{\Phi}}
\def\bPs{\bm{\Psi}}
\def\bphi{\bm{\phi}}
\def\bvph{\bm{\varphi}}
\def\brho{\bm{\rho}}
\def\hG{\widehat{\mathbf {\Gamma}}}
\def\vv{{\bf{v}}}
\def\w{{\bf{w}}}

\begin{document}

\title{Two-Scale Approach to an Asymptotic Solution of Maxwell Equations in Layered Periodic Medium} %Title of paper

\author{M. V. Perel}
\email{m.perel@spbu.ru} \affiliation{St. Petersburg State University, 7/9 Universitetskaya nab., St. Petersburg 199034, Russia. }

\author{M. S. Sidorenko}
\email{m-sidorenko@yandex.ru} \affiliation{St. Petersburg State University, 7/9 Universitetskaya nab., St. Petersburg 199034, Russia. }

\date{\today}

\begin{abstract}
An  asymptotic investigation of  monochromatic electromagnetic fields in a layered periodic
medium is carried out in the assumption that the wave frequency is close to the frequency of a stationary
point of a dispersion surface.
%The stationary points are common for the both TM and TE polarizations.
We find solutions of Maxwell equations  by the method of two-scale asymptotic expansions.
 %Scales characterize  a period of the medium and the variation of the envelopes of the field.
We establish that the principal order of the expansion of a solution dependent on three spatial
coordinates is the sum of two differently polarized Floquet-Bloch solutions, each of which is
multiplied by a slowly varying envelope function. We derive that the envelope functions satisfy a
system of differential equations with constant coefficients. In new variables, it is reduced to a system of two independent equations, both of them are
either  hyperbolic or  elliptic, depending on the type of the stationary point.  The envelope functions are
independent only in the planar case. Some consequences are discussed.
\end{abstract}

\maketitle %\maketitle must follow title, authors, abstract and \pacs

\section{Introduction}

The propagation of electromagnetic waves in media with periodic changes of dielectric permittivity
and magnetic permeability, in the so-called photonic crystals, is a subject of many investigations
(see, for example, \cite{Joannopoulos}). The popularity of such studies is caused, on the one hand,
by unusual properties of such media, which are yet not completely known and which promise new
applications.
% such as controlling of the light propagation.
On the other hand, there is a technological progress in the creation of artificial structures with
prescribed properties.

The problem of wave propagation in a quickly oscillating medium can be usually reduced to a
problem of wave propagation in an effective homogeneous medium. The mathematic methods applied in
finding the effective medium are methods of homogenization \cite{BahvalovPanasenko}, \cite{Jikov},
\cite{Palencia}, i.e.,  asymptotic methods in the long wavelength approximation:  $kb \to
\infty$, where $k$ is the wavenumber, $b$ is the period of oscillations of the medium.  One of the methods
of the derivation of asymptotic formulas is the  method of two-scale expansions. The solution is
assumed to  depend on  fast and slow distance variables. It turns out that the principal term  is a
function of only a slow variable and satisfies  homogenized equations, which may be
interpreted as equations in an effective medium. Another approach is based on the spectral point
of view: the field can be represented as a superposition of  Floquet-Bloch solutions
corresponding to  the lower part of the  spectrum.

A lot of physical works use the concept of  effective medium. Peculiarities of a
dispersive surface in an  effective medium are responsible for unusual phenomena of wave propagation in
such structures, which  are often not observed in nature. For example, if one of the principal
components of the effective electric or magnetic tensor has an  opposite sign with respect to other two
principal components, the dispersive surface in such a medium has a hyperbolic point.  Such media are
named  hyperbolic ones and are studied intensively; see, for example, \cite{Poddubny}. Another
example is a composite material consisting of layers of metals and dielectrics.  The
isofrequency dispersive surface for such a structure may have a plane part, i.e., one of the
components of the wavevector may be almost independent of the other two. The waves may propagate
without distortion in such a medium, see \cite{Belov}.

The present work was motivated by  papers of  Longhi %of  S. Longhi and D. Janner
\cite{Longhi}, \cite{LonghiJanner}, in which the possibilities of the existence  of localized waves in
2D and 3D periodic structures were studied. The idea was to take the frequency of a monochromatic
electromagnetic field equal to the frequency of the stationary point of the dispersive (band)
surface in the periodic structure.  The field in 2D and 3D photonic crystals was represented as a
superposition of  Wannier functions with an envelope, which satisfies the equation with
constant coefficients. For hyperbolic (saddle) point  the envelope satisfies the wave equation,
where  one of the spatial coordinates stands for  time. By choosing for the envelope one of the
known solutions of the wave equation with finite energy, see, for example, \cite{LocalWav}, one may
obtain localized waves  in the periodic structure. The papers \cite{Longhi}, \cite{LonghiJanner}
were mainly concentrated on physical aspects of the problem. We are more interested  in  careful
mathematical analysis and obtain new results in the 3D case.

We noted in our paper \cite{PerelSidorenko2012} that the local hyperbolic behavior of a
dispersive surface occurs even for the simplest dielectric  layered periodic structures, for
example, for a  structure with  alternating layers of dielectric. In this case we studied  a
field dependent only on two spatial coordinates in \cite{PerelSidorenko2012} and found and
numerically confirmed the following physical phenomenon: undistorted beams can propagate in such a structure and there are
only two permitted directions of a beam in the medium, which differs by a sign from the angle with
the normal to layers. A similar phenomenon was found before by numerical simulation in 2D crystals in
\cite{ChienTang};  but it  was not investigated analytically.

In the present paper, we study monochromatic electromagnetic fields with a frequency close to that of a
stationary point of the dispersive surface, which may be a hyperbolic or an elliptic one.  The
field is studied in a layered structure with periodically varying dielectric permittivity and the
magnetic permeability. We deal  with fields dependent on all the three spatial coordinates.  Our
aim is to elaborate a mathematical asymptotic approach for the description of  solutions
of  Maxwell equations.

The two-scale method in the homogenization theory is applied usually to equations written in
divergent form. Our asymptotic scheme is applied to  Maxwell equations in  matrix form, which
was first suggested in the book by  Felsen and Marcuvitz \cite{Felsen}. This form was also used in
the turning points problem for the Maxwell equations in \cite{PerelRadiophys},  for other
particular equations in \cite{PerelFialkovskiiKiselev}, \cite{PerelKaplunovRogerson},
\cite{AndronovZaikaPerel}, and for the  operators in the general form  \cite{FialkovskyPerel2016unifying}.  Therefore we believe that results may have generalizations to other
problems. The matrix form of the Maxwell equations is given in Section \ref{sec:matrix}.

In Section \ref{S_disp_rel} we discuss  difficulties in studying electromagnetic fields dependent
on three spatial variables  caused by the fact that the stationary points are common for waves of
both the TM and the TE polarizations, and  that the concepts of TM and TE polarizations depend on
the direction of propagation and  at the stationary point itself  they loose their meaning. In Section
\ref{sec:help}, we obtain some integral formulas for second derivatives of the dispersion functions,
which are applied in the next section. In Section \ref{two-scale},  we develop the two-scale
expansions method for our problem. Apart from the period $b$ in $z$ direction we introduce the second
scale of length, which is a scale of field variation in the  $(x,y)$ plane denoted by $L$. We regard
the ratio between $b$ and  $L$ as a small parameter $\chi$. We consider the entire formal
asymptotic series and show that  the recurrent system for subsequent approximations can be solved
step by step. In Section \ref{two-scale}, we found that the principal term of the two-scaled field
is obtained as a sum of two differently polarized Floquet-Bloch solutions at the stationary point
and each of these solutions has a slowly varying envelope function. The envelope functions satisfy
a system of two partial differential equations with constant coefficients.
  In Section
\ref{sec:alpha}, we show that these equations are independent only if the envelopes depend on two spatial coordinates. A  qualitative consequence of equations is that undistorted beams can propagate in the medium.
In the general 3D case,
each envelope function can be expressed in terms of two functions. These functions  are solutions of
two equations, which are of the Helmholtz type or the Klein-Gordon-Fock type  for elliptic
or hyperbolic stationary points, respectively.  The fact that the wave field is described by a system of two equations was not predicted in the  papers  \cite{Longhi}, \cite{LonghiJanner}.
  Our paper is concluded with two Appendices. Appendix 1 contains known results important for the present paper  from the Floquet-Bloch theory, which  are given in our notation. Appendix 2 contains proofs of two  lemmas.

Our case may be regarded as  homogenization near a stationary point of the dispersive surface in
contrast to a more usual situation, where the frequency corresponds to the lower part of the spectrum
and obtained in the assumption that $kb \to 0$.  Our case demands an assumption $k b \sim
1/\sqrt{\varepsilon_{av} \mu_{av}}$, where $\varepsilon_{av}$, $\mu_{av}$ are typical permittivity
and permeability.
%  i.e., permittivity and permeability in the medium are of order of
%$\varepsilon_{av}$, $\mu_{av},$ respectively.
The equations for  envelopes are analogous to equations in an effective medium and can be used for
a qualitative description of the wavefield.

\section{Statement of the problem}

A monochromatic electromagnetic field satisfies the Maxwell
equations
\begin{equation}\label{eq}
\begin{array}{ccc}
{\rm rot}\mathbb{E} = i k \mu \mathbb{H},   \\
{\rm rot}\mathbb{H} = -i k \varepsilon \mathbb{E},
\end{array}
\end{equation}
where $\varepsilon(z+b) = \varepsilon(z)$, \quad $\mu(z+b) = \mu(z)$; $\varepsilon$ and $\mu$ are
piecewise continuous.

The medium with alternating dielectric layers is a practically important special case.

%We assume that the variables $\mathbb{E}$ and $\mathbb{H}$ and the parameters $\varepsilon$ and
%$\mu$ in formulas (\ref{eq}) are already normalized as follows
%\begin{equation}\nonumber
%\mathbb{E} =
%\sqrt{\frac{\varepsilon_{av}}{\mu_{av}}}\widetilde{\mathbb{E}},
%\quad \mathbb{H} =
%\sqrt{\frac{\varepsilon_{av}}{\mu_{av}}}\widetilde{\mathbb{H}},
%\quad {\rm where} \quad \varepsilon=
%\frac{\widetilde{\varepsilon}}{\varepsilon_{av}}, \quad \mu =
%\frac{\widetilde{\mu}}{\mu_{av}}, \quad k =
%\sqrt{\varepsilon_{av}\mu_{av}} \widetilde{k},
%\end{equation}
%where $\varepsilon_{av}$, $\mu_{av}$ are typical dielectric permittivity and magnetic permeability,
%these parameters may be large, $\varepsilon$ and $\mu$ are of order unity, and
%$\widetilde{\varepsilon}, \widetilde{\mu}$ are the original parameters of the equation. The
%variables ${\widetilde{k}}$ and $k$ mean the wave number in vacuum and in the medium with
%parameters $\varepsilon_{av}$ and $\mu_{av}$, respectively.

We seek solutions under two assumptions. The first assumption is about the parameters of the
problem. The Maxwell equations contain two parameters of length dimension: the wavelength $\lambda
= 2 \pi/k$ and  the period of the medium $b$. We introduce the third parameter of length dimension
$L$, which is the scale of variation of the field in the $(x,y)$ plane.  {We assume that the
parameter $\chi \equiv b/L$ is small:}
\begin{equation}
\chi = b/L \ll 1.
\end{equation}
The second assumption is related to the frequency of the monochromatic field $\omega$. To state
this assumption, we need some concepts: the quasimomentum $\vp$, the dispersive surface
$\omega=\omega(\vp)$, and Floquet-Bloch solutions. We determine and discuss in detail all these
concepts in Appendix 1. This assumption means that the frequency $\omega$ is close to that of the stationary
point of the dispersive surface $\omega_*$:
\begin{equation}
\omega = \omega_* +\chi^2 \delta\omega, \qquad  {\delta \omega \sim 1,}
\end{equation}
where $\omega_*$  satisfies the condition
\begin{equation}\label{stat-point}
\left. \nabla \omega \right|_{\vp_*} =0, \quad
\omega_*=\omega(\vp_*).
\end{equation}
We shall see that these stationary points are minima and saddle
points. There is also an additional condition, which we shall
introduce later.

\section{Matrix form of  Maxwell equations \label{sec:matrix} }

We  represent the Maxwell equations in matrix form for the sake of
brevity and generality of subsequent asymptotic considerations:
\begin{align}
& k P \bPs  = - i \widehat{ \G}\cdot\nabla \bPs, \quad \bPs = \left(
\begin{array}{c} \mathbb{E} \\ \mathbb{H} \end{array} \right), \label{Maxwell} \\
& \hG\cdot\nabla \equiv\G_1 \frac{\partial}{\partial x} + \G_2
\frac{\partial}{\partial y} + \G_3 \frac{\partial}{\partial z},
\qquad \label{scal-def}
\end{align}
where
\begin{equation}\label{P-G1-G2}
\begin{aligned}
\P = \left( \begin{array}{cc} \varepsilon I & 0 \\ 0 & \mu  I \\
\end{array} \right), \,\, \G_1 = \left( \begin{array}{cc}
0 & \gamma_1 \\ -\gamma_1 & 0 \\
\end{array} \right), \,\, \G_2 = -\left( \begin{array}{cc}
0 & \gamma_2 \\ -\gamma_2 & 0 \\
\end{array} \right), \,\, \G_3 = \left( \begin{array}{cc}
0 & \gamma_3 \\ -\gamma_3 & 0 \\
\end{array} \right),
\nonumber
\end{aligned}
\end{equation}
\begin{equation}\label{g2-g3}
\begin{aligned}
\gamma_1 = \left(\begin{array}{ccc} 0 & 0 & 0 \\ 0 & 0 & 1 \\ 0 & -1
& 0 \\ \end{array} \right), \quad \gamma_2 =
\left(\begin{array}{ccc} 0 & 0 & 1 \\ 0 & 0 & 0 \\ -1 & 0 & 0 \\
\end{array} \right), \quad \gamma_3 =
\left(\begin{array}{ccc} 0 & 1 & 0 \\ -1 & 0 & 0 \\ 0 & 0 & 0 \\
\end{array} \right),
\end{aligned}
\end{equation}
and $k/\sqrt{\varepsilon_{av} \mu_{av}} = \omega/c$, where $c$ is
the speed of light in vacuum.

%Our aim is to obtain asymptotics of solutions of the Maxwell equations in the following form
%\begin{equation}\label{Psi-form}
%\bPs = \bPs(z, \brho), \quad \xi=\chi x, \quad \eta = \chi y, \quad
%\zeta = \chi z, \quad \brho=(\xi,\eta,\zeta)
%\end{equation}
%if $\chi \ll 1.$ There are two scales of the field in the vertical direction:  a slow coordinate
%$\zeta= \chi z$ and the coordinate $z$. The field in the lateral directions $x$ and $y$ depends
%only on the slow variables $\xi=\chi x$ and  $\eta = \chi y$.

We introduce two types of  inner products. Let ${\bf v}(z)$ and ${\bf w}(z)$ be 6-component
complex-valued vector functions; then
\begin{equation}\label{scal-angle}
<{\bf v}, {\bf w}> = \sum\limits_{j=1}^6 \overline{v^j} w^j, \qquad
j=1,\ldots 6,
\end{equation}
where the bar over the symbol stands for complex conjugation. If
${\bf v}$ and ${\bf w}$ are periodic with  period $b$ and piecewise
continuous, we  define $({\bf v},{\bf w})$ as follows:
\begin{equation}\label{scalar-int}
({\bf v},{\bf w}) = \int\limits_0^b <{\bf v}(z), {\bf w}(z)> dz.
\end{equation}
The Umov-Poynting vector, averaged over time, for a monochromatic
field of  frequency $\omega$, i.e., the energy flux density of this
field, averaged over $T=2\pi/\omega$, is determined as
\begin{equation}\label{def-Sk}
\vec{s} = \frac{1}{2} \Re{\, \overline{\mathbb{E}}\times
{\mathbb{H}}}.
\end{equation}
It is easy to check that
\begin{equation}\label{Point}
<{\bPs},\G_j {\bPs}> = 4s_j, \quad j=1,2,3; \quad ({\bPs},\G_j {\bPs}) =  {2} \int\limits_0^b
\Re{\, \left[\overline{\mathbb{E}} \times {\mathbb{H}}\right]}_j dz.
\end{equation}
The density of electromagnetic energy, averaged over time, in the
case of real $\varepsilon$ and $\mu$ reads
\begin{equation}\label{def-u}
u = \frac{\varepsilon}{4} |\mathbb{E}|^2 + \frac{\mu}{4}
|\mathbb{H}|^2.
\end{equation}
We note that
\begin{equation}\label{def-en}
<{\bPs},\P {\bPs}> = 4 u, \qquad ({\bPs},\P {\bPs}) =
\int\limits_0^b ( \varepsilon |\mathbb{E}|^2 + \mu |\mathbb{H}|^2)
dz.
\end{equation}
%The round brackets denotes the integral of the square brackets over the period of the medium.
% This means that we first average over the time period and then over the spatial period.

\section{The Floquet-Bloch solutions and the dispersion relation \label{S_disp_rel}}

Since the properties of the medium do not depend on $x,y$, we shall seek  {particular}
solutions in the form
\begin{equation}\label{plane_wave}
\bPs_B(x, y, z;\vp) = e^{i(p_x x + p_y y)} \bPh(z;\vec{p}), \quad
\bPh = \left( \begin{array}{c} {\mathbf E} \\ {\mathbf{H}}
\end{array} \right),
\end{equation}
where the parameters $p_x$ and $p_y$ are lateral components of the wave vector. The components of
the vector-valued function $\bPh(z;\vec{p})$ satisfy a system of ordinary differential equations
with periodic coefficients:
\begin{equation}\label{Maxwell-pxpy}
k\mathbf{P} \bPh + i\G_3\frac{\partial \bPh}{\partial z} = p_x\G_1 \bPh + p_y\G_2 \bPh.
\end{equation}
We are going to obtain Floquet-Bloch solutions of this system. However there are difficulties owing
to the vector nature of the problem. It is well known (see, for example, \cite{Felsen}) that an
 appropriate choice of the coordinate system allows one to split the system
(\ref{Maxwell-pxpy}) into two independent subsystems. These subsystems describe waves of two
polarizations: the transverse electric wave and the transverse magnetic waves, which are named the
TE and TM waves, respectively. We call the coordinates, in which such a splitting occurs, the natural
ones.

\subsection{Two types of  Floquet-Bloch solutions in the natural coordinates}

To determine the type of a  {solution}, we should clarify, which component of the wave is
transverse to the propagation plane, the electric or the magnetic one. The propagation plane passes
through the lateral wave vector $(p_x,p_y)$ and the $z$ axis. To find the TM and TE modes, it is
convenient to rotate the axes in the $(x,y)$ plane through an angle $\gamma$, so that in the new
coordinates, $\widetilde{p_y} = 0$, $\widetilde{p_x} \equiv p_{\parallel} = \sqrt{p_x^2 + p_y^2}$.
We denote fields in the rotated coordinate system by a tilde. The waves of TE and TM types are as follows:
\begin{equation}\label{TE-TM}
{\widetilde  {\bPh}}^{E} = \left( \begin{array}{cccccc} 0,
E_{\perp}, 0, H_{\parallel}, 0, \frac{p_{\parallel}}{k\mu} E_{\perp}
\end{array} \right)^t,\quad {\widetilde {\bPh}}^{H} =
\left(\begin{array}{cccccc} E_{\parallel}, 0,
-\frac{p_{\parallel}}{k\varepsilon} H_{\perp}, 0, H_{\perp}, 0
\end{array} \right)^t.
\end{equation}
 The system of equations (\ref{Maxwell-pxpy}) in such coordinates splits into two subsystems:
\begin{equation}\label{Maxwell-TM}
\left\{\begin{array}{rcl} \displaystyle{ i\frac{\partial
{H}_{\perp}}{\partial z}} & = &  - k\varepsilon {E}_{\parallel},  \\
\\ \displaystyle{  i\frac{\partial {E}_{\parallel}}{\partial z}} & = &
-\left(\frac{k^2\varepsilon\mu  -
p_{\parallel}^2}{k\varepsilon}\right) {H}_{\perp}.
\end{array} \right., \quad \left\{\begin{array}{rcl} \displaystyle{
i\frac{\partial {H}_{\parallel}}{\partial z}} & = &
\left(\frac{k^2\varepsilon\mu - p^2_{\parallel}}{k\mu}\right)
{E}_{\perp},  \\ \\ \displaystyle{  i\frac{\partial
{E}_{\perp}}{\partial z}} & = & k\mu {H}_{\parallel}.
\end{array} \right.
\end{equation}
In the special case  $p_{\parallel}=0$, the  splitting into TM and TE waves has no meaning:
both systems can be reduced to the following one:
\begin{equation}\label{E0-H0}
i \frac{dE_0}{dz} = - k \mu H_0;\quad i \frac{d H_0}{dz} = - k
\varepsilon E_0,
\end{equation}
where
\begin{equation}\label{limEH-perp-par}
{E}_{\parallel}|_{p_{\parallel}=0} = E_0, \quad  {H}_{\perp}|_{p_{\parallel}=0} = H_0, \quad {E}_{\perp}|_{p_{\parallel}=0} = -E_0, \quad {H}_{\parallel}|_{p_{\parallel}=0} = H_0.
\end{equation}
We introduce the new notation for vector-functions obtained by means of the passage to the limit  $p_{\parallel}\to 0$
in ${\widetilde \bPh}^{E}$,${\widetilde \bPh}^{H}$:
\begin{equation}
{\widetilde  {\bPh}}^{E} |_{p_{\parallel} \to 0} \to \quad \bPh^X, \quad {\widetilde  {\bPh}}^{H}
|_{p_{\parallel} \to 0} \to \quad \bPh^Y,
\end{equation}
where
\begin{equation}\label{Bas-XY}
\bPh^X=(E_0,0,0,0,H_0,0)^t, \quad \bPh^Y=(0,-E_0,0,H_0,0,0)^t.
\end{equation}
The superscript $X$ (or $Y$) indicates that the vector-function has the  nonzero first (or second) component.

The obtained systems of ordinary linear differential equations (\ref{Maxwell-TM}), (\ref{E0-H0})
for ${E}_{\parallel}$, ${H}_{\perp}$  and ${E}_{\perp}$,${H}_{\parallel}$ and for $E_0$ and $H_0$,
respectively, are systems with piecewise continuous  periodic coefficients, because $\varepsilon$ and
$\mu$ are piecewise continuous. Each system has two Floquet-Bloch solutions;  for details, see
Appendix 1.  We denote the components of the second solution by the subscript $2$. For
example, the  {two} solutions of the first subsystem of (\ref{Maxwell-TM}) for TM waves read
\begin{equation}\label{Floquet-TE}
 {\left(\begin{array}{c} E_{\parallel}\\
H_{\perp} \end{array}\right) = e^{ i p_z z} {\bf
U}^H_{+}(z;p_z,p_{\parallel}^2,\omega), \quad \left(\begin{array}{c} E_{\parallel\, 2}\\
H_{\perp \,2} \end{array}\right) = e^{-i p_z z} {\bf U}^H_{-}(z;p_z,p_{\parallel}^2,\omega).}
\end{equation}
 The solutions depend on the parameters of the equations $p_{\parallel}^2$ and $\omega$, and
also on the real-valued parameter $p_z$, which is called the quasimomentum and which is related to
$p_{\parallel}^2$ and $\omega$ by the formula
\begin{equation}\label{disp-p-o}
 p_z = p_z(p_{\parallel}^2,\omega ).
\end{equation}
% Formula (\ref{disp-p-o}) arises from the periodicity conditions
%\begin{equation}
%{\bf U}^E(z+b;p_{\parallel}^2,\omega)={\bf U}^E(z;p_{\parallel}^2,\omega),
%\end{equation}
%and from the continuity of the tangential components of $\vec{E}$
%and $\vec{H}$.
 This relation ensures that the functions ${\bf U}^H_{\pm}$ are periodic,
\begin{equation}
{\bf U}^H_{\pm}(z+b;p_z,p_{\parallel}^2,\omega)={\bf U}^H_{\pm}(z;p_z,p_{\parallel}^2,\omega),
\end{equation}
and continuous as  functions of the $z$ variable. We call these functions {\it Floquet-Bloch
amplitudes}. The relation (\ref{disp-p-o}) can be reduced to the following one:
\begin{equation}\label{disp-relH}
\omega=\omega^H(\vp).
\end{equation}
We call this equation  the {\it dispersion relation}. The function $\omega^H(\vp)$ is called the {\it
dispersion function}. It is a multisheeted function on $[-\pi/b,\pi/b)\times \mathbb{R}^2_+$, and
its derivation is discussed in the Appendix 1. Substituting (\ref{disp-relH}) into (\ref{Floquet-TE}), we obtain $E_{\parallel}$ and $H_{\perp}$ as functions of $z$ and $\vp$.
  The Floquet-Bloch solutions and the dispersion
function for the waves of TE type $({E}_{\perp}$, ${H}_{\parallel})$ and for the solutions $E_0$ and
$H_0$ of the system (\ref{E0-H0}) are obtained analogously. The solutions (\ref{Floquet-TE}) are
linearly independent if $p_z \ne 0, \pm \pi/b$; see Appendix 1. However, this particular case $p_z = 0, \pm \pi/b$ is important in the present paper.

%where $p_z$ is the quasi-momentum, which is connected with $p_{\parallel}$ and $\omega$ by the
%dispersion equation, which arises from the periodicity conditions
%\begin{equation}
%{\bf U}^E(z+b;p_{\parallel}^2,\omega)={\bf U}^E(z;p_{\parallel}^2,\omega),
%\end{equation}
%and from the continuity of the tangential components of $\vec{E}$
%and $\vec{H}$. We denote the dispersion relation by $\omega=\omega(\vp)$, which is a multivalued
%function on $\mathbb{R}^2\times[-\pi/b,\pi/b)$; its derivation is in the Appendix 1. In fact the
%function $\omega=\omega_j(\vp)$ depends on $p_x,p_y$ through $\sqrt{p_x^2+p_y^2}$.

We are interested here in effects that arise if the frequency of the problem is
close to the frequency of one of the stationary points of some sheet of the dispersive surface. In
 Appendix 1, we show that the  multisheeted dispersive surfaces $\omega = \omega^H(\vp)$
 {and $\omega = \omega^E(\vp)$} have stationary points on each sheet, at the points $\vp_*$,
$p_{\parallel*} = 0$, $p_{z*} = 0, \pm\pi/b$, where $\nabla\omega^E_{\vp*}=0$.  In these
points the sheets of the dispersive surfaces for TM and TE polarizations touch each other, and
therefore $\nabla\omega^H_{\vp_*}=\nabla\omega^E_{\vp_*}=0$ and $\omega^E(\vp_*)=\omega^H(\vp_*) \equiv
\omega_*$. At the point $\vp_*$,  formulas (\ref{Floquet-TE}) may give the same solution, which is
bounded at infinity and which is periodic for $p_{z*} = 0$ and anti-periodic for $p_{z*} =
\pm\pi/b.$ Then the second solution, which is linearly independent with the first one, grows at infinity
( for details, see Appendix 1). It is this case that is of interest to us in the present paper. The solutions read
\begin{equation}\label{two_sol2}
\begin{aligned}
\left.\left(\begin{array}{c} E_{\parallel}\\
H_{\perp} \end{array}\right)\right|_{\vp=\vp_*}  = \left.\left(\begin{array}{c} E_{0}\\
H_{0} \end{array}\right)\right|_{p_{z}=p_{z*}}\, & = &
e^{ip_{z*}z} \vec{U}^H_+(z;p_{z*},0,\omega_*),  \qquad \qquad \qquad \qquad\\
\left.\left(\begin{array}{c} E_{\parallel\, 2}\\
H_{\perp \,2} \end{array}\right)\right|_{\vp=\vp_*}  = \left.\left(\begin{array}{c} E_{02}\\
H_{02} \end{array}\right)\right|_{p_{z}=p_{z*}} & = & e^{ip_{z*}z} \left[e^{-ip_{z*}b}
\frac{z}{b} \vec{U}^H_+(z;p_{z_*},0,
\omega_*) + \vec{Q}^H(z;p_{z_*}, \omega_*) \right],
\end{aligned}
\end{equation}
where the vector-valued function ${\bf Q}^H$ is a periodic and continuous function of the
variable $z$. The function ${\bf Q}^H$ does not depend on the argument $p_{\parallel*}^2,$ because
$p_{\parallel*}^2=0,$ and we are not going to use the second solution (\ref{two_sol2}) for $p_{\parallel}\ne0$.

The solutions of the second subsystem (\ref{Maxwell-TM}) are expressed in the following
way:
\begin{equation}
\left.\left(\begin{array}{c} E_{\perp}\\
H_{\parallel} \end{array}\right)\right|_{\vp=\vp_*}\, = \left.\left(\begin{array}{c} -E_{0}\\
\quad H_{0} \end{array}\right)\right|_{p_{z}=p_{z*}}, \quad \left.\left(\begin{array}{c} E_{\perp \, 2}\\
H_{\parallel \,2} \end{array}\right)\right|_{\vp=\vp_*}  = \left.\left(\begin{array}{c} - E_{02}\\
\quad H_{02} \end{array}\right)\right|_{p_{z}=p_{z*}}.
\end{equation}

\subsection{Floquet-Bloch solutions in an arbitrary coordinate system}

We have found solutions $E_{\parallel},
H_{\perp}$ and $E_{\perp}, H_{\parallel}$  of the systems (\ref{Maxwell-TM}). These solutions were found in the coordinate system
rotated by an angle $\gamma$, which characterizes the direction of wave propagation. In the initial
coordinate system,  these solutions read:
\begin{equation}\label{Phi_E_Phi_H}
{{\bPh}}^H  = \left(\begin{array}{c} E_{\parallel} \cos \gamma \\
E_{\parallel} \sin\gamma \\ -\frac{1}{k \varepsilon}p_{\parallel}
H_{\perp} \\ -H_{\perp} \sin\gamma
\\ H_{\perp} \cos\gamma \\ 0
\end{array} \right), \quad
{{\bPh}}^E = \left(\begin{array}{c} -E_{\perp} \sin\gamma \\
E_{\perp} \cos \gamma \\ 0 \\ H_{\parallel} \cos\gamma
\\ H_{\parallel} \sin\gamma \\ \frac{1}{k \mu}p_{\parallel} E_{\perp}
\end{array} \right).
\end{equation}
The solutions (\ref{Phi_E_Phi_H}) do not have a limit for $p_{\parallel} \to 0$ as a function of two
variables $p_x,p_y$. This limit  exists in any direction determined by $\gamma$
and depends on the direction $\gamma$:
\begin{equation}\label{lim-old}
{\bPh}^H |_{{p_{\parallel}}\to 0} \to \cos{\gamma} \bPh^X -
\sin{\gamma} \bPh^Y,\quad {\bPh}^E|_{{p_{\parallel}}\to 0} \to
\sin{\gamma} \bPh^X + \cos{\gamma} \bPh^Y.
\end{equation}
Here we take into account (\ref{limEH-perp-par}) and the definitions of $\bPh^X$ and $\bPh^Y$ from (\ref{Bas-XY}).
%In particular, if $\gamma=0$, then $p_{\parallel}=p_x$. If
%$\gamma=\pi/2$, then $p_{\parallel}=p_y$. We obtain from (\ref{lim})
%\begin{eqnarray}
%&&{\bPh}^H |_{p_x\to 0,p_y=0} \to \quad \bPh^X,\qquad
%{\bPh}^E|_{p_x\to 0, p_y=0} \to  \bPh^Y, \label{lim-p1}\\
%&&{\bPh}^H |_{p_x=0, p_y\to 0} \to - \bPh^Y,\qquad {\bPh}^E|_{p_x=0,
%p_y\to 0} \to  \bPh^X.\label{lim-p2}
%\end{eqnarray}

 Below we will use the derivatives $\partial {{\bPh}}^H / \partial p_j$, $\partial
{{\bPh}}^E/\partial p_j,$ $j=x,y,z$, which are  linear combinations of the derivatives of the
functions $E_{\parallel}, H_{\perp}$ and $E_{\perp}, H_{\parallel}$. These functions are
analytic functions of the parameter $p_{\parallel}^2$, since the coefficients of the system
(\ref{Maxwell-TM}) are  analytic functions of $p_{\parallel}^2$. It is easy to obtain
\begin{equation}
\frac{\partial E_{\parallel}}{\partial p_x} = 2 p_{\parallel} \frac{\partial
E_{\parallel}}{\partial p_{\parallel}^2} \cos\gamma.
\end{equation}
This yields  $\partial E_{\parallel}/\partial p_x |_{p_{\parallel}=0}=0.$ Similarly, we
derive that the derivatives of $H_{\perp}$, $E_{\perp}$ and $H_{\parallel}$ with respect to the
variables $p_x$ and $p_y$ are zero for $p_{\parallel}=0$. Finally, we get
\begin{eqnarray}
&& \frac{\partial {\bPh}^H}{\partial p_x}\left|_{p_x=p_y=0}\right. = \frac{\partial
{\bPh}^H}{\partial p_y}\left|_{p_x=p_y=0}\right.
= \left(0,0,-\frac{H_0}{k\varepsilon},0,0,0\right)^t , \label{der-p1}\\
&& \frac{\partial {\bPh}^E}{\partial p_x}\left|_{p_x=p_y=0}\right. = \frac{\partial
{\bPh}^E}{\partial p_y}\left|_{p_x=p_y=0}\right. = \left(0,0,0,0,0,-\frac{E_0}{k\mu}\right)^t.
\label{der-p2}
\end{eqnarray}
These derivatives do not depend on the direction $\gamma$.

We introduce the unified notation $\bPh^f$ for the Floquet-Bloch solutions of different types:
\begin{equation}\label{Bl-sol-amp}
\bPh^f{(z;\vp)} = e^{ip_zz}\bvph^f(z;\vp), \quad f=H, E, X, \, {\rm or} \, Y,
\end{equation}
where $\bvph^f(z+b;\vp)=\bvph^f(z;\vp)$.
The six-component vector-functions $\bvph^f(z,\vp)$  are
called  the Floquet-Bloch amplitudes.
Let us discuss  formula (\ref{Bl-sol-amp}) in more detail.
 The Floquet-Bloch solutions $\bPh^f{(z;\vp)}$ for
$f=H,E$ are determined by  formula (\ref{Phi_E_Phi_H}), where $E_{\parallel}, H_{\perp}$ and
$E_{\perp}, H_{\parallel}$ are the first solutions (\ref{Floquet-TE}) of the systems
(\ref{Maxwell-TM}). We assume that  the frequency $\omega$  in the formulas (\ref{Floquet-TE}) is expressed in terms of $\vp$ on one  of the sheets of the multisheeted function $\omega=\omega^f(\vp)$. We  also take into account the fact that $p_{\parallel}^2 = p_x^2 + p_y^2$, $\cos{\gamma}=p_x/p_{\parallel},$ $\sin{\gamma} = p_y/p_{\parallel}$. The  number of the sheet is omitted for brevity. The Floquet-Bloch solutions $\bPh^f(z;\vp)$, $f=X,Y$,
are defined only for $p_{\parallel}=0$, i.e., for $\vp = \vp_0 \equiv (0,0,p_z)$; their
polarization is not defined.

The limit of the Floquet-Bloch amplitudes in the direction determined by $\gamma$ follows from
(\ref{lim-old}):
\begin{equation}\label{lim}
{\bvph}^H |_{{p_{\parallel}}\to 0} \to \cos{\gamma} \bvph^X - \sin{\gamma} \bvph^Y,\quad
{\bvph}^E|_{{p_{\parallel}}\to 0} \to \sin{\gamma} \bvph^X + \cos{\gamma} \bvph^Y.
\end{equation}
In particular, if $\gamma=0$, then $p_{\parallel}=p_x$. If $\gamma=\pi/2$, then
$p_{\parallel}=p_y$.  From (\ref{lim}) we obtain
\begin{eqnarray}
&&{\bvph}^H{(z;\vp)} |_{p_x\to 0,p_y=0} \to \quad \bvph^X{(z;\vp_0)},\quad
{\bvph}^E{(z;\vp)}|_{p_x\to 0, p_y=0} \to  \bvph^Y{(z;\vp_0)}, \label{lim-p1}\\
&&{\bvph}^H{(z;\vp)}|_{p_x=0, p_y\to 0} \to - \bvph^Y{(z;\vp_0)},\quad {\bvph}^E{(z;\vp)}|_{p_x=0,
p_y\to 0} \to  \bvph^X{(z;\vp_0)}.\label{lim-p2}
\end{eqnarray}

Next, we need to obtain the derivatives of the solutions $\bvph^H$ and $\bvph^E$ with respect to
the parameters $p_x$ and $p_y$ at the point $ p_{\parallel}= 0$. Owing  to (\ref{der-p1})
and(\ref{der-p2}) and the definition (\ref{Bl-sol-amp}), we get
\begin{eqnarray}
&& \frac{\partial {\bvph}^H}{\partial p_x}\left|_{\vp_*}\right. = \frac{\partial
{\bvph}^H}{\partial p_y}\left|_{\vp_*}\right.
= \left.\left(0,0,-\frac{H_0}{k\varepsilon},0,0,0\right)^t\right|_{p_{z*}} e^{-ip_{z*}z} , \label{der-p1v}\\
&& \frac{\partial {\bvph}^E}{\partial p_x}\left|_{\vp_*}\right. = \frac{\partial
{\bvph}^E}{\partial p_y}\left|_{\vp_*}\right. =
\left.\left(0,0,0,0,0,-\frac{E_0}{k\mu}\right)^t\right|_{p_{z*}} e^{-ip_{z*}z}. \label{der-p2v}
\end{eqnarray}
These derivatives do not depend on the direction $\gamma$. These
functions are periodic in $z$, because ${\bvph}^H$, ${\bvph}^E$ are
periodic  for all $\vp$.

\section{Some auxiliary relations \label{sec:help}}

Our aim is to obtain relations for the derivatives of the dispersion  {functions} by means of
the Floquet-Bloch amplitudes. To do this, we give equations for these
amplitudes and their derivatives with respect to the parameters $p_j$, $j=x,y,z$.

\subsection{Equations for the Floquet-Bloch amplitudes and their derivatives }

 We deal with six-component solutions
of the Maxwell equations:
\begin{equation}\label{sol-Bl-am}
\bPs^f_B {(x,y,z;\vp)} = e^{i(p_xx+p_yy)}\bPh^f {(z;\vp)},  \quad f=E, H,
\end{equation}
where $\bPh^f$ is of the form (\ref{Bl-sol-amp}).
Here the superscript $f$ stands for the type of the field: $f=E$ corresponds to the TE polarization
and $f=H$ corresponds to the TM one. Inserting (\ref{sol-Bl-am}) into Maxwell equations, we rewrite
them in the form
\begin{equation}\label{Af}
\A^f(\vp)\bvph^f {(z;\vp)} =0,
\end{equation}
where
\begin{equation}
\A^f(\vp) \equiv
\frac{\omega^f(\vp)}{c}
\P + i \G_3\frac{\partial}{\partial z} - \vp \cdot \hG, \qquad
\vp \cdot \hG \equiv p_x\G_1 + p_y \G_2 + p_z \G_3. \label{Bl-amp2}
\end{equation}
We note that the equations for both types differ only by the
dispersion function $\omega^f(\vp)$. These equations are valid for
any $\vec{p}$, and thus we can take the derivatives of these
equations with respect to $\vec{p}$.
By taking the derivatives with respect to parameters $p_j, j=x,y$, we obtain
\begin{eqnarray}
&&\A^f(\vp) \frac{\partial \bvph^f}{\partial p_j} =
-  \frac{\partial \A^f}{\partial p_j} \bvph^f, \label{eq-der}\\
&&\A^f(\vp) \frac{\partial^2 \bvph^f}{\partial p_j^2} = -
\frac{1}{c} \frac{\partial^2 \omega^f}{\partial p_j^2} \P
\bvph^f - 2 \frac{\partial \A^f(\vp)}{\partial p_j}
\frac{\partial \bvph^f}{\partial p_j},  \label{eq-der-2}
\end{eqnarray}
where
\begin{equation}
\frac{\partial \A^f}{\partial p_j} =  \frac{1}{c}
\frac{\partial \omega^f}{\partial p_j} \P - \G_j. \label{der-A}
\end{equation}
 The derivatives with respect to $p_z$ are discussed below at the end of the Section. Now we proceed with specifying
formulas (\ref{eq-der}), (\ref{der-A}) at the stationary points $\vp_*$. Such points for each
branch of the multisheeted dispersion functions $\omega^f(\vp)$, $f=E,H$, differ only by $p_{z*}$, and all of them
have $p_{\parallel*}=0$.  Solutions $\bvph^f(z;\vp)$ are not continuous
as functions of two variables $p_x$ and $p_y$ near the point $p_x=p_y=0$.  However, for any fixed
angle $\gamma$, they can be calculated by means of the passage to the limit,  see (\ref{lim}). The derivatives at $p_{\parallel}=0$ do not depend on $\gamma$, see (\ref{der-p1v}), (\ref{der-p2v}). We
indicate these derivatives at $\vp=\vp_*$ by the asterisk subscript. Upon substitution  {$\vp=\vp_*$}, the operator $\A$
%takes the form
 is denoted as
\begin{equation}\label{A-star}
 \A^f(\vp_*)\equiv \A_*(\vp_*), \quad f = E, H,
\end{equation}
and no more depends on the wave type TM or TE.

 Let $\mathcal{M}$ be a class of six-component vector-valued functions of $z$, which are periodic with a period $b$,
  piecewise  smooth on the period and their components with numbers  $1,2,4,5$ are continuous; see details in Appendix 2.
  Lemma 1 (see Appendix 2) shows that the
operator $\A_*(\vp_*)$ is symmetric on the functions from $\mathcal{M}$.

Finally,  passing to the limit in (\ref{eq-der}), (\ref{der-A}), in view of (\ref{lim-p1}),
(\ref{lim-p2}) we obtain
\begin{eqnarray}
&&\A_* \frac{\partial \bvph^H_*}{\partial p_x} = \G_1 \bvph^X_*,
\qquad \A_* \frac{\partial \bvph^E_*}{\partial p_x} =
\G_1 \bvph^Y_*, \label{der-1-0}\\
&&\A_* \frac{\partial {\bvph}^H_*}{\partial p_y} =  - \G_2
{\bvph}^Y_*, \qquad \, \A_* \frac{\partial {\bvph}^E_*}{\partial
p_y} =  \G_2 {\bvph}^X_*. \label{der-2-0}
\end{eqnarray}
 Here we have introduced the notation
\begin{equation}\label{vph-XY-def}
\bvph^X(z;\vp_*)=\bvph^X_*, \quad \bvph^Y(z;\vp_*)=\bvph^Y_*.
\end{equation}
We note that the terms containing the derivative $\partial \omega^f/\partial p_j$ vanish at the
stationary point, since this derivative vanishes.

For the second derivatives of the Floquet-Bloch amplitudes, in view of  (\ref{eq-der-2}),
(\ref{der-A}), and  (\ref{lim-p1}), (\ref{lim-p2}) we derive
\begin{eqnarray}
&&\A_* \frac{\partial^2 \bvph^H_*}{\partial p_x^2} = - \frac{1}{c} \frac{\partial^2
\omega^H_*}{\partial p_x^2}
\P \bvph^X_* + 2 \G_1 \frac{\partial \bvph^H_*}{\partial p_x}, \label{der-2H1}\\
&&\A_* \frac{\partial^2 \bvph^E_*}{\partial p_x^2} = - \frac{1}{c} \frac{\partial^2
\omega^E_*}{\partial p_x^2}
\P \bvph^Y_* + 2 \G_1 \frac{\partial \bvph^E_*}{\partial p_x},\label{der-2E1}\\
&&\A_* \frac{\partial^2 \bvph^H_*}{\partial p_y^2} =  \frac{1}{c} \frac{\partial^2
\omega^H_*}{\partial p_y^2}
\P \bvph^Y_* + 2 \G_2 \frac{\partial \bvph^H_*}{\partial p_y}, \label{der-2H2} \\
&&\A_* \frac{\partial^2 \bvph^E_*}{\partial p_y^2} = - \frac{1}{c} \frac{\partial^2
\omega^E_*}{\partial p_y^2} \P \bvph^X_* + 2 \G_2 \frac{\partial \bvph^E_*}{\partial
p_y}.\label{der-2E2}
\end{eqnarray}
 We emphasize that the directional limit of ${\bvph}^H$ and ${\bvph}^E$ at the point $\vp_*$ can be
expressed by ${\bvph}^X_*$, as well as by ${\bvph}^Y_*$ depending on the direction $\gamma$.

Now we proceed to the calculation of the derivatives with respect to $p_z$ at the point $\vp_*$, i.e., $p_{\parallel*}=0$, $p_z=p_{z*}$.
  If $p_{\parallel}= p_{\parallel*}=0$, i.e., $\vp = \vp_0 \equiv
(0,0,p_z)$,  we have
\begin{equation}\label{om0}
\omega^H(\vp_0)=\omega^E(\vp_0)\equiv \omega^0(\vp_0)
\end{equation}
and $\A^H(\vp_0)=\A^E(\vp_0)\equiv \A^0(\vec{p}_0)$. The Floquet-Bloch amplitudes
$\bvph^X$, $\bvph^Y$  (see (\ref{Bl-sol-amp}), (\ref{Bas-XY}) )
 satisfy  the same equation
\begin{equation}\label{A0}
\A^0(\vp_0) \bvph^f(z;\vp_0) = 0, \qquad f = X, \,\, Y.
\end{equation}
By  differentiating  (\ref{A0}) with respect to $p_z$ and then by passing to the limit $p_z \to
p_{z*}$, we get
\begin{eqnarray}
&&\A_* \frac{\partial {\bvph}^X_*}{\partial p_z} = \G_3 {\bvph}^X_*,
\quad \A_* \frac{\partial {\bvph}^Y_*}{\partial p_z} = \G_3
{\bvph}^Y_*. \label{der-3-0}
\end{eqnarray}
 By taking the second derivatives of (\ref{A0}) and then by passing to the limit $p_z \to p_{z*}$, we
obtain
\begin{eqnarray}
&&\A_* \frac{\partial^2 \bvph^X_*}{\partial p_z^2} = - \frac{1}{c}
\frac{\partial^2 \omega^0_*}{\partial p_z^2} \P {\bvph}^X_* +
2 \G_3\frac{\partial \bvph^X_*}{\partial p_z},\label{der-2X3} \\
&&\A_* \frac{\partial^2 \bvph^Y_*}{\partial p_z^2} = - \frac{1}{c}
\frac{\partial^2 \omega^0_*}{\partial p_z^2} \P {\bvph}_*^Y +
2 \G_3\frac{\partial \bvph^Y_*}{\partial p_z}.\label{der-2Y3}
\end{eqnarray}

\subsection{Derivatives of dispersion functions}

Now we get integral relations containing derivatives of the Floquet-Bloch amplitudes and
derivatives of dispersion functions. We take the inner product of (\ref{der-1-0}), (\ref{der-2-0}),
and (\ref{der-3-0}) with $\bvph^X_*$ or $\bvph^Y_*$ and, taking into account the fact that
\begin{equation}
\A_* \bvph^X_* = 0, \quad \A_* \bvph^Y_* = 0,
\end{equation}
by Lemma 1 (see Appendix 2) we find
\begin{equation}\label{matr-el1}
\left(\bvph^{f_2}_*,\G_j \bvph^{f_1}_*\right)= 0 \quad {\rm
for\quad any}\quad j=1,2,3; \quad f_1=X,Y; \quad f_2=X,Y.
\end{equation}
Henceforth, we use the fact that $\bvph^{f_2}$ and its derivatives  with respect to parameters $p_j$, $j=x,y,z$  belong to $\mathcal{M}$.

We are going to apply the same operations to formulas  {(\ref{der-2H1}--\ref{der-2E2}) and
(\ref{der-2X3}--\ref{der-2Y3})}. We introduce the notation
\begin{equation}\label{u_def}
u_*^{f_1f_2}\equiv\left(\bvph_*^{f_1},\P \bvph_*^{f_2}\right),
\quad f_1=X \,\, \mathrm{or} \,\, Y, \quad f_2=X \,\, \mathrm{or} \,\,Y,
\end{equation}
where $u^{ff}$ has the meaning of the  density of energy averaged over time (see Section \ref{sec:matrix}), and
\begin{equation}
\ddot{\omega}^f_{11*} \equiv \left. \frac{\partial^2 \omega^f}{\partial p_x^2}
\right|_{\vp=\vp*}, \quad \ddot{\omega}^f_{22*} \equiv \left. \frac{\partial^2
\omega^f}{\partial p_y^2} \right|_{\vp=\vp*},  \quad f=H \,\, \mathrm{or} \,\,
E; \quad \ddot{\omega}^0_{33*} \equiv \left. \frac{\partial^2 \omega^0}{\partial
p_z^2} \right|_{\vp=\vp*},\label{omega_dot_dot}
\end{equation}
where $\ddot{\omega}^0_{33*}=\ddot{\omega}^H_{33*}=\ddot{\omega}^E_{33*}$, because (\ref{om0}) is
valid for any $p_z$. We arrive at the relations
\begin{eqnarray}
&&\frac{\ddot{\omega}^H_{11*}}{c}  u_*^{XX} = 2\left(\bvph_*^X,\G_1
\frac{\partial \bvph^H_*}{\partial p_x}\right), \quad
\frac{\ddot{\omega}^E_{11*}}{c}  u_*^{YY} = 2\left(\bvph_*^Y,\G_1
\frac{\partial \bvph^E_*}{\partial p_x}\right), \label{om2-1}\\
&&\frac{\ddot{\omega}^H_{22*}}{c}  u_*^{YY} = -2\left(\bvph_*^Y,\G_2
\frac{\partial \bvph^H_*}{\partial p_y}\right), \quad
\frac{\ddot{\omega}^E_{22*}}{c}  u_*^{XX} = 2\left(\bvph_*^X,\G_2
\frac{\partial \bvph^E_*}{\partial p_y}\right), \label{om2-2}\\
&&\frac{\ddot{\omega}^0_{33*}}{c}  u_*^{XX} =2\left(\bvph_*^X,\G_3
\frac{\partial \bvph^X_*}{\partial p_z}\right), \quad
\frac{\ddot{\omega}^0_{33*}}{c}  u_*^{YY} = 2\left(\bvph_*^Y,\G_3
\frac{\partial \bvph^Y_*}{\partial p_z}\right). \label{om2-3}
\end{eqnarray}
If $f_1 \neq f_2$, then $u^{f_1f_2}\equiv 0$. Moreover, $u^{XX} = u^{YY}$ by the definitions of  $\bvph_*^X$, $\bvph_*^Y$, and $\P$.

\subsection{Additional relations}
Multiplying equations (\ref{der-1-0}), (\ref{der-2-0}) and
(\ref{der-3-0}) by $\bvph_*^X$ and $\bvph_*^Y$ in such a way that on
the left-hand side we obtain $(\bvph_*^X,\P\bvph_*^Y)$, which
vanishes, we find the following relations:
\begin{eqnarray}
&&\left(\bvph_*^Y,\G_1 \frac{\partial \bvph^H_*}{\partial
p_x}\right)= \left(\bvph_*^X,\G_1 \frac{\partial \bvph^E_*}{\partial
p_x}\right)=
\left(\bvph_*^Y,\G_2 \frac{\partial \bvph^H_*}{\partial p_y}\right)=0, \label{help1}\\
&&\left(\bvph_*^X,\G_2 \frac{\partial \bvph^E_*}{\partial
p_y}\right)=\left(\bvph_*^Y,\G_3 \frac{\partial \bvph^X_*}{\partial
p_z}\right)=\left(\bvph_*^X,\G_3 \frac{\partial \bvph^Y_*}{\partial
p_z}\right)=0. \label{help2}
\end{eqnarray}

Now we mention some other useful relations with the derivatives of
$\bvph^f$. The derivatives of $\bvph^f$ with respect to $p_x$ and
$p_y$ coincide; see (\ref{der-p1}) and (\ref{der-p2}). This fact
yields
\begin{equation}\label{eq-G-bv}
\G_1 \bvph^X_* = - \G_2 {\bvph}^Y_*, \quad \G_1 \bvph^Y_* = \G_2
{\bvph}^X_*.
\end{equation}
The same relations can be obtained by direct computations by (\ref{Bas-XY}), (\ref{P-G1-G2}) not
only at the point $\vp_*$ but at the point $\vp=(0,0,p_z)$ for any $p_z$. The direct calculations
with the help of (\ref{der-p1}), (\ref{der-p2}) and (\ref{P-G1-G2}) show that
\begin{equation}\label{g3d}
\G_3 \frac{\partial {\bvph}^H_*}{\partial p_j}=0, \quad \G_3
\frac{\partial {\bvph}^E_*}{\partial p_j}=0, \quad j=x,y.
\end{equation}

\section{The two-scaled asymptotic decomposition  \label{two-scale}}

We give an asymptotic representation of some special solutions of Maxwell equations in the entire
space under several assumptions:
\begin{enumerate}
\item the vertical period  {$b$} of the medium is small as compared with the horizontal scale
of the field, and the relation between the scales is characterized by the small parameter $\chi$,
\item the frequency $\omega$ is close to the  {frequency $\omega_*$ of the stationary point $\vp_*$ of one
of the sheets of the dispersion function $\omega=\omega^f(\vp),$ $f=H,E$, i.e., the frequency
$\omega_*$ is determined by the relations
\begin{equation}
\omega_*=\omega^E(\vp_*)=\omega^H(\vp_*),\quad \nabla \omega^f(\vp_*)=0,\quad f=H,E.
\end{equation}
We assume that
\begin{equation}
\omega = \omega_* + \chi^2 \delta\omega, \quad \delta\omega \sim 1.
\end{equation}}
%stationary point $(\vp_*, \omega_*)$ of the dispersion functions $\omega=\omega(\vp)$, i.e.,
%$\omega = \omega_* +\chi^2 \delta\omega$. At the point $\vp_*$, there must be one bounded and one
%unbounded solutions of the periodic problem (\ref{E0-H0}).
\item  {We assume that  there is one bounded and one unbounded Floquet-Bloch solution of the
periodic problem (\ref{E0-H0}) at the point $\vp_*$.}
\end{enumerate}
 {Our aim is to find the asymptotics of  solutions of the Maxwell equations in the following
form:
\begin{equation}\label{Psi-form}
\bPs = \bPs(z, \brho), \quad \xi\equiv\chi x, \quad \eta \equiv \chi y, \quad \zeta \equiv \chi z,
\quad \brho=(\xi,\eta,\zeta)
\end{equation}
where $\chi \ll 1.$  In the direction transverse to the layers, the field  has two scales, one of them is
determined by the slow variable $\zeta= \chi z$, and the other is given by the variable $z$. In the plane of the layers in the directions $x$ and $y$, the field depends only on the slow variables $\xi=\chi x$, $\eta
= \chi y$.}

We seek  a solution in the form of a two-scaled asymptotic series
\begin{eqnarray} \label{anz}
&&\bPs(z,\brho) = \bPh(z,\brho) e^{i (p_{x*}\xi +
p_{y*}\eta)/{\chi}}, \quad
\bPh(z,\brho) = \bphi(z,\brho) e^{i p_{z*} z},\\
&&\bphi(z,\brho) = \sum\limits_{n \geq 0} \chi^n
\bphi^{(n)}(z,\brho), \quad \bphi^{(n)}(z + b,\brho) =
\bphi^{(n)}(z,\brho), \quad \brho = (\xi,\eta,\zeta).
\end{eqnarray}
In the case under
consideration, the stationary point is $p_{x*}=p_{y*}=0$.

We assume that $\bphi^{(n)} \in \mathcal{M}$ for every $n$ as functions of $z$, see the definition after formula (\ref{A-star}).  Also these functions are  infinitely
differentiable with respect to slow variables.

The Maxwell equations in new variables read
\begin{equation}\label{Maxw}
k_* \P \bPs + i\G_3\frac{\partial\Psi}{\partial z} = -i\chi
\widehat{\G}\cdot \nabla_{\brho} \bPs - \chi^2 \frac{\delta
\omega}{c} \P \bPs, \quad k_* = \frac{\omega_*}{c},
\end{equation}
where
\begin{equation}
\widehat{\G}\cdot \nabla_{\brho}  \equiv \G_1 \frac{\partial}{\partial
\xi} + \G_2 \frac{\partial}{\partial \eta} + \G_3
\frac{\partial}{\partial \zeta}. \nonumber
\end{equation}

Substituting the asymptotic series (\ref{anz}) in (\ref{Maxw}),  we
obtain a set of equations
\begin{equation}
\A_* \bphi^{(0)} = 0, \qquad \A_* \bphi^{(n)} = \vec{F}^{(n)}, \label{Maxwell-series}
\end{equation}
where
\begin{eqnarray}
&&\A_*\bPs =  k_* \P \bPs + i\G_3\frac{\partial\bPs}{\partial z} - p_{3*} \G_3 \bPs,\label{def-A*}\\
&&\vec{F}^{(1)} = -i \widehat{\G}\cdot \nabla_{\rho}\bphi^{(0)},\\
&&\vec{F}^{(n)} = -i \widehat{\G}\cdot \nabla_{\rho}\bphi^{(n-1)} -
\frac{\delta \omega}{c} \P \bphi^{(n-2)},\quad n \ge2. \label{Fn}
\end{eqnarray}

 Prior to solving the set of equations, we are going to find the relations between the
parameters of the problem, which ensures that all nonzero terms on the right-hand side of
(\ref{def-A*}) are of the same order. We assume that the variables $\mathbb{E}$ and $\mathbb{H}$
and the parameters $\varepsilon$ and $\mu$ in formulas (\ref{eq}) are already normalized as follows
\begin{equation}\nonumber
\mathbb{E} = \sqrt{\frac{\varepsilon_{av}}{\mu_{av}}}\widetilde{\mathbb{E}}, \quad \mathbb{H} =
\sqrt{\frac{\varepsilon_{av}}{\mu_{av}}}\widetilde{\mathbb{H}}, \quad {\rm where} \quad
\varepsilon= \frac{\widetilde{\varepsilon}}{\varepsilon_{av}}, \quad \mu =
\frac{\widetilde{\mu}}{\mu_{av}}, \quad k_* = \sqrt{\varepsilon_{av}\mu_{av}} \widetilde{k_*},
\end{equation}
where $\varepsilon_{av}$, $\mu_{av}$ are typical dielectric permittivity and magnetic permeability,
these parameters may be large, $\varepsilon$ and $\mu$ are of order unity, and
$\widetilde{\varepsilon}, \widetilde{\mu}$ are the original parameters of the equation. The
variables ${\widetilde{k}}_*$ and $k_*$ mean the wave number in vacuum and in the medium with
parameters $\varepsilon_{av}$ and $\mu_{av}$, respectively,  {$k_* = \sqrt{\varepsilon_{av}
\mu_{av}}\omega/c $, where $c$ is the speed of light in vacuum. The second and the third (if
nonzero) terms in the right-hand side of (\ref{def-A*}) are of  order of $1/b$. The
first and the second terms are of the same order if
\begin{equation}
\sqrt{\varepsilon_{av} \mu_{av}}\omega/c \sim 1/b.
\end{equation}
This means that the case under consideration differs from the well-known case  $\omega b/c \to
0$.}

\subsection{The principal order}
The equation of  principal order  term is the equation for the Floquet-Bloch amplitudes at
the stationary point $\vp_*$.  It does not contain the derivatives with respect to slow
variables $\brho=(\xi, \eta, \zeta)$ and its coefficients do not depend on $\brho$. Its solutions
may depend on $\brho$ as on parameters. We seek the principal term in the form of
%It depends on $\brho=(\xi, \eta, \zeta)$ as on a vector of parameters. Its solutions read
\begin{equation}\label{princip}
\bphi^{(0)}(z, \brho) = \alpha_{1}(\brho)\bvph^{X}_*(z) +
\alpha_{2}(\brho) \bvph^{Y}_*(z), \quad \bphi^{(0)} \in \mathcal{M},
\end{equation}
where
\begin{equation}
\bvph^f_*(z) = \bvph^f(z;\vp_*) = e^{-ip_{z*}z} \bPh^f{(z;\vp_*)},
\end{equation}
$f = X,\,\,Y$, and $\bPh^f{(z;\vp)}$ for $p_{\parallel}=0$ are defined in (\ref{Bas-XY}) by means of the functions $E_0$ and $H_0$, which satisfy a system (\ref{E0-H0}). The
functions $\alpha_{1}, \alpha_{2}$ are  arbitrary scalar functions of  slow variables $\brho$.
Additional restrictions on these arbitrary functions will arise later.

\subsection{First-order approximation}

The equation for the first-order  term $\bphi^{(1)}$ of the expansion has the form
\begin{equation}
\A_* \bphi^{(1)} = -i\hG\cdot\nabla_{\brho} \bphi^{(0)}
\label{Maxwell-1-order}, \quad \bphi^{(1)} \in \mathcal{M}.
\end{equation}
   In order to get the solution of the system belonging to the class $\mathcal{M}$, we must impose  additional conditions.

%This system is nonhomogeneous, so we must prove that the solution exists prior to solving it.

\textbf{Lemma 2}. A solution from the class $\mathcal{M}$ of the equation $\A_* \bphi = \vec{F}$  exists if and only if the following conditions are satisfied:
\begin{equation}\label{lemma2}
\left(\bvph^X_*,\vec{F}\right)=0,\quad
\left(\bvph^Y_*,\vec{F}\right)=0.
\end{equation}
The proof of the Lemma 2 is given in the Appendix 2.

Now we check the solvability conditions for the first-order
approximation (\ref{Maxwell-1-order}), i.e., we must check that
\begin{equation}
\left(\bvph^X_*,\hG\cdot\nabla_{\brho} \bphi^{(0)}\right)=0,\quad
\left(\bvph^Y_*,\hG\cdot\nabla_{\brho} \bphi^{(0)}\right)=0,
\end{equation}
which are reduced to the following conditions
\begin{eqnarray}
&&\left(\bvph^X_*,\hG\bvph^X_*\right)\cdot \nabla_{\brho}\alpha_1 +
\left(\bvph^X_*,\hG\bvph^Y_*\right)\cdot \nabla_{\brho}\alpha_2 =0,\\
&&\left(\bvph^Y_*,\hG\bvph^X_*\right)\cdot \nabla_{\brho}\alpha_1 +
\left(\bvph^Y_*,\hG\bvph^Y_*\right)\cdot \nabla_{\brho}\alpha_2 =0,
\end{eqnarray}
where, for example,
\begin{equation}
\left(\bvph^X_*,\hG\bvph^X_*\right)\cdot \nabla_{\brho}\alpha_1 \equiv \left(\bvph^X_*,
\G_1\bvph^X_*\right) \frac{\partial \alpha_1}{\partial \xi} + \left(\bvph^X_*, \G_2\bvph^X_*\right)
\frac{\partial \alpha_1}{\partial \eta} + \left(\bvph^X_*, \G_3\bvph^X_*\right) \frac{\partial
\alpha_1}{\partial \zeta}.
\end{equation}
These conditions are satisfied at the stationary point owing to
(\ref{matr-el1}).

Now let us find the exact formula for the solution $\bphi^{(1)}$. The right-hand side  of the
equation (\ref{Maxwell-1-order}) can be written as follows:
\begin{equation}\label{ord1}
\vec{F}^{(1)} = -i \hG\cdot (\nabla_{\brho}\alpha_1) \bvph^X_* -i
\hG\cdot (\nabla_{\brho}\alpha_2 )\bvph^Y_*.
\end{equation}
Taking into account (\ref{eq-G-bv}), we replace the terms containing $\G_2$ by the terms containing
$\G_1$. Collecting the resulting terms, we obtain
\begin{equation}
\vec{F}^{(1)} = -i \left( \frac{\partial \alpha_1}{\partial \xi} -
\frac{\partial \alpha_2}{\partial \eta}\right) \G_1 \bvph^X_* -i
\left( \frac{\partial \alpha_1}{\partial \eta} + \frac{\partial
\alpha_2}{\partial \xi}\right) \G_1 \bvph^Y_*  -i \frac{\partial
\alpha_1}{\partial \zeta} \G_3  \bvph^X_* - i\frac{\partial
\alpha_2}{\partial \zeta} \G_3 \bvph^Y_*.
\end{equation}
Now the right-hand side contains four terms. Instead of solving (\ref{Maxwell-1-order}), we solve
four independent vector equations with right-hand sides  containing each of the terms and then take
the sum of their solutions. We add also a solution of the homogeneous equation. The particular
solutions of four equations coincide with solutions of (\ref{der-1-0}), (\ref{der-2-0}), and
(\ref{der-3-0}). We find the following solution:
\begin{eqnarray}
\label{Phi_1_solution} &&\bphi^{(1)} = -i \left( \frac{\partial
\alpha_1}{\partial \xi} - \frac{\partial \alpha_2}{\partial
\eta}\right) \frac{\partial \bvph^H_*}{\partial p_x} -i \left(
\frac{\partial \alpha_1}{\partial \eta} + \frac{\partial
\alpha_2}{\partial \xi}\right) \frac{\partial \bvph^E_*}{\partial
p_x}
\nonumber\\
&& -i \frac{\partial \alpha_1}{\partial \zeta} \frac{\partial
\bvph^X_*}{\partial p_z} - i\frac{\partial \alpha_2}{\partial \zeta}
\frac{\partial \bvph^Y_*}{\partial p_z}
 + \alpha_{1}^{(1)}\bvph^{X}_*(z) + \alpha_{2}^{(1)}
\bvph^{Y}_*(z),
\end{eqnarray}
where $\alpha_{1,2}^{(1)}$ are  new arbitrary functions of the slow variables
$\brho=(\xi,\eta,\zeta)$. The subscript $*$ stands to show that all the derivatives with the
respect to $p_j, j=x,y,z$ are taken at $\vp=\vp_*$.

\subsection{The second-order approximation\label{sec_second_approx}}
Now let us consider the equation  {on the second-order approximation}
\begin{equation}
\A_* \bphi^{(2)} = -i\hG \cdot\nabla_{\brho}\bphi^{(1)}, \quad
\bphi^{(2)} \in \mathcal{M}. \label{Maxwell-2-order}
\end{equation}
The solution $\bphi^{(1)}$ depends on four unknown functions
$\alpha_{j}, j=1,2$ and $\alpha^{(1)}_{j}, j=1,2.$ The solvability
conditions of (\ref{Maxwell-2-order}) yield equations for two of
them:
\begin{eqnarray}
&&i \left(\bvph^X_*,\hG\cdot\nabla_{\brho} \bphi^{(1)}\right) +
\left(\bvph^X_*,\frac{\delta \omega}{c} \P \bphi^{(0)}\right)=0, \label{cond-2X}\\
&& i \left(\bvph^Y_*,\hG\cdot\nabla_{\brho} \bphi^{(1)}\right) +
\left(\bvph^Y_*,\frac{\delta \omega}{c} \P \bphi^{(0)}\right)=0.
\label{cond-2Y}
\end{eqnarray}
For brevity, we introduce the notation
\begin{equation}\label{tau-def}
\tau_1 \equiv \left( \frac{\partial \alpha_1}{\partial \xi} -
\frac{\partial \alpha_2}{\partial \eta}\right), \quad \tau_2 \equiv
\left( \frac{\partial \alpha_1}{\partial \eta} + \frac{\partial
\alpha_2}{\partial \xi}\right).
\end{equation}
Let us calculate every term separately
\begin{eqnarray}\label{cond2X-0}
i\left(\bvph^X_*,\hG\cdot\nabla_{\brho} \bphi^{(1)}\right) =
\left(\bvph^X_*,\hG \frac{\partial
\bvph^H_*}{\partial p_x}\right) \cdot \nabla_{\brho}\tau_1  + \left(\bvph^X,\hG
\frac{\partial \bvph^E_*}{\partial p_x}\right) \cdot \nabla_{\brho}\tau_2
\nonumber \\
+ \left(\bvph^X_*,\hG \frac{\partial \bvph^X_*}{\partial p_z}\right)\cdot \nabla_{\brho}
\frac{\partial \alpha_1}{\partial \zeta} + \left(\bvph^X_*,\hG \frac{\partial \bvph^Y_*}{\partial
p_z}\right) \cdot \nabla_{\brho} \frac{\partial \alpha_2}{\partial \zeta}.
\end{eqnarray}
 Here we take into account that the coefficients of $\alpha_{2}^{(1)}$ and $\alpha_{1}^{(1)}$
vanish owing to (\ref{matr-el1}).

For example, the first term in the right-hand side of (\ref{cond2X-0}) reads
\begin{eqnarray}
\left(\bvph^X_*,\hG \frac{\partial
\bvph^H_*}{\partial p_x}\right) &\cdot& \nabla_{\brho}\tau_1 \equiv \nonumber\\
 & &\left(\bvph^X_*,\G_1 \frac{\partial \bvph^H_*}{\partial
p_x}\right) \frac{\partial \tau_1}{\partial
\xi} + \left(\bvph^X_*,\G_2 \frac{\partial \bvph^H_*}{\partial
p_x}\right)\frac{\partial \tau_1}{\partial
\eta} + \left(\bvph^X_*,\G_3 \frac{\partial \bvph^H_*}{\partial
p_x}\right) \frac{\partial \tau_1}{\partial
\zeta}.\label{cond2X-01}
\end{eqnarray}
According to (\ref{om2-1}), the first term is proportional to $\ddot{\omega}_{11*}^H$. The second
term vanishes owing to (\ref{der-p1}) and the last relation of (\ref{help1}), the third term
vanishes by (\ref{g3d}). Analogously,
\begin{eqnarray}
\left(\bvph^X_*,\hG \frac{\partial
\bvph^E_*}{\partial p_x}\right) &\cdot& \nabla_{\brho}\tau_2  \equiv \nonumber\\
& & \left(\bvph^X_*,\G_1
\frac{\partial \bvph^E_*}{\partial p_x}\right)\frac{\partial \tau_2}{\partial
\xi}  + \left(\bvph^X_*,\G_2 \frac{\partial \bvph^E_*}{\partial
p_x}\right)\frac{\partial \tau_2}{\partial
\eta} + \left(\bvph^X_*,\G_3 \frac{\partial \bvph^E_*}{\partial
p_x}\right) \frac{\partial \tau_2}{\partial
\zeta}.
\label{cond2X-02}
\end{eqnarray}

According to (\ref{om2-2}), the second term is proportional to $\ddot{\omega}_{22*}^E$ and two
other terms vanish: the first one owing to (\ref{help1}), and the third one because of (\ref{g3d}).

Now we proceed to the last two terms in (\ref{cond2X-0}). By (\ref{om2-3})-(\ref{help2}), we obtain
\begin{eqnarray}
&& \left(\bvph^X_*,\widehat{\G} \frac{\partial \bvph^X_*}{\partial
p_z}\right) \cdot \nabla_{\brho}\frac{\partial \alpha_1}{\partial \zeta} = \left(\bvph^X_*,\G_3
\frac{\partial \bvph^X_*}{\partial p_z}\right)\frac{\partial^2 \alpha_1}{\partial \zeta^2} =
\frac{ \ddot{\omega}^0_{33*}}{2c} u_*^{XX} \frac{\partial^2 \alpha_1}{\partial \zeta^2}, \label{cond2X-03}\\
&& \left(\bvph^X_*,\widehat{\G} \frac{\partial \bvph^Y_*}{\partial
p_z}\right) \cdot \nabla_{\brho}\frac{\partial \alpha_2}{\partial \zeta} = \left(\bvph^X_*,
\G_3 \frac{\partial \bvph^Y_*}{\partial p_z}\right) \frac{\partial^2 \alpha_2}{\partial \zeta^2}= 0. \label{cond2X-04}
\end{eqnarray}

We note that since $u_*^{XY}=0$, we get, with account of (\ref{princip}) and (\ref{u_def}),
\begin{equation}
\left(\bvph^X_*,\frac{\delta \omega}{c} \P \bphi^{(0)}\right)= \frac{\delta \omega}{c}u_*^{XX}
\alpha_1.
\end{equation}
Finally, the condition (\ref{cond-2X}) yields
\begin{equation}\label{eq0-1}
\frac{\partial \tau_1}{\partial \xi} \ddot{\omega}_{11*}^H +
\frac{\partial \tau_2}{\partial \eta} \ddot{\omega}_{22*}^E +
\frac{\partial^2 \alpha_1}{\partial \zeta^2}\ddot{\omega}_{33*}^0 +
2\frac{\delta \omega}{c}\alpha_1=0.
\end{equation}
We omit here the nonzero common factor $u_*^{XX}/2=u_*^{YY}/2$.

Now we obtain the second equation by considering the condition
(\ref{cond-2Y}). We use a similar line of argument. First, we take
the relation that differs from (\ref{cond2X-0}) only by the first
factor in  inner products
\begin{eqnarray}
i\left(\bvph^Y_*,\hG\cdot\nabla_{\brho} \bphi^{(1)}\right) =
\left(\bvph^Y_*,\hG \frac{\partial
\bvph^H_*}{\partial p_x}\right) \cdot \nabla_{\brho}\tau_1 + \left(\bvph^Y,\hG
\frac{\partial \bvph^E_*}{\partial p_x}\right) \cdot \nabla_{\brho}\tau_2  \nonumber\\
+ \left(\bvph^Y_*,\hG \frac{\partial \bvph^X_*}{\partial
p_z}\right) \cdot \nabla_{\brho} \frac{\partial \alpha_1}{\partial \zeta} +
\left(\bvph^Y_*,\hG \frac{\partial \bvph^Y_*}{\partial p_z}\right) \cdot
\nabla_{\brho} \frac{\partial \alpha_2}{\partial \zeta}.
\label{cond2Y-0}
\end{eqnarray}
Here, for example,
\begin{eqnarray}
\left(\bvph^Y_*,\hG \frac{\partial
\bvph^H_*}{\partial p_x}\right) &\cdot& \nabla_{\brho} \tau_1  \equiv \nonumber \\
& & \left(\bvph^Y_*,
\G_1 \frac{\partial \bvph^H_*}{\partial p_x}\right)\frac{\partial \tau_1}{\partial
\xi} + \left(\bvph^Y_*,\G_2 \frac{\partial \bvph^H_*}{\partial
p_x}\right) \frac{\partial \tau_1}{\partial
\eta} + \left(\bvph^Y_*,\G_3 \frac{\partial \bvph^H_*}{\partial
p_x}\right) \frac{\partial \tau_1}{\partial
\zeta}.
\end{eqnarray}
The second term here is proportional to $\ddot{\omega}^H_{22_*}$ according to (\ref{om2-2}). The
first and the third terms vanish by (\ref{help1}) and (\ref{g3d}), respectively. The term
$\left(\bvph^Y_*,\widehat{\G} \left.\partial \bvph^E_*\right/\partial
p_x\right)\cdot\nabla_{\brho}\tau_2$ is treated analogously to (\ref{cond2X-02}) by using the
second relation of (\ref{om2-1}), (\ref{help2}) with (\ref{der-2-0}), and (\ref{g3d}). Analogously
to (\ref{cond2X-03}) and (\ref{cond2X-04}), we obtain
\begin{equation}
\left(\bvph^Y_*,\widehat{\G} \frac{\partial \bvph^X_*}{\partial p_z}\right)\cdot \nabla_{\brho}
\frac{\partial \alpha_1}{\partial \zeta} = 0, \quad \left(\bvph^Y_*,\widehat{\G} \frac{\partial
\bvph^Y_*}{\partial p_z}\right) \cdot\nabla_{\brho}\frac{\partial \alpha_2}{\partial \zeta} =
\frac{\ddot{\omega}^0_{33*}}{2c}u_*^{XX} \frac{\partial^2 \alpha_2}{\partial \zeta^2}  .
\end{equation}
\begin{equation}
\left(\bvph^Y_*,\hG \frac{\partial \bvph^H_*}{\partial p_x}\right) \cdot\nabla_{\brho}
\tau_1=-\frac{\ddot{\omega}^H_{22_*}} {2c} u_*^{XX}\frac{\partial \tau_1}{\partial
\eta}.\label{cond2Y-01}
\end{equation}

Again, by (\ref{help1}) and (\ref{help2}), we get
\begin{equation}\label{eq0-2}
- \frac{\partial \tau_1}{\partial \eta} \ddot{\omega}_{22*}^H + \frac{\partial \tau_2}{\partial
\xi} \ddot{\omega}_{11*}^E + \frac{\partial^2 \alpha_2}{\partial \zeta^2}\ddot{\omega}_{33*}^0 +
2\frac{\delta \omega}{c}\alpha_2=0.
\end{equation}

Taking into account the definition of $\tau_1$ and $\tau_2$ (\ref{tau-def}) and the fact that
$\ddot{\omega}_{11*}^H=\ddot{\omega}_{22*}^H$ and $\ddot{\omega}_{11*}^E=\ddot{\omega}_{22*}^E$, we
rewrite the equations for $\alpha_1$ and $\alpha_2$ as follows:
\begin{equation}\label{eq0-sys}
\begin{aligned}
\frac{\partial^2 \alpha_1}{\partial \xi^2} \ddot{\omega}_{11*}^H +
\frac{\partial^2 \alpha_1}{\partial \eta^2} \ddot{\omega}_{11*}^E +
\frac{\partial^2 \alpha_1}{\partial \zeta^2}\ddot{\omega}_{33*}^0 +
2\frac{\delta \omega}{c}\alpha_1- \frac{\partial^2
\alpha_2}{\partial \xi \partial \eta}
(\ddot{\omega}_{11}^H-\ddot{\omega}_{11}^E)=0, \\
\frac{\partial^2 \alpha_2}{\partial \xi^2} \ddot{\omega}_{11*}^E +
\frac{\partial^2 \alpha_2}{\partial \eta^2} \ddot{\omega}_{11*}^H +
\frac{\partial^2 \alpha_2}{\partial \zeta^2}\ddot{\omega}_{33*}^0 +
2\frac{\delta \omega}{c}\alpha_2-\frac{\partial^2 \alpha_1}{\partial
\xi \partial \eta}(\ddot{\omega}_{11}^H-\ddot{\omega}_{11}^E)=0.
\end{aligned}
\end{equation}

\subsection{Higher-order approximations}

 Now we  turn to the set of equations
(\ref{Maxwell-series}-\ref{Fn}). By considering several recurrent equations, we conclude that the approximation of $n$th order has the form
\begin{eqnarray}\label{sol-phi-n}
&&\bphi^{(n)} =\alpha_{1}^{(n)}\bvph^{X}_*(z) + \alpha_{2}^{(n)}\bvph^{Y}_*(z) +
 {G^{(n)}\left(\alpha_1^{(n-1)}, \alpha_2^{(n-1)},  \ldots \alpha_1, \alpha_2 \right)},
\end{eqnarray}
 {where $G^{(n)}$ is the linear combination of the derivatives  of the
functions $\alpha_j^{(k)}, j = 1,2, \, k = 1 \ldots n-1$, with respect to the variables
$\xi,\eta,\zeta$ with known coefficients. For example, in the principal order, owing  to the formula
(\ref{princip}), $G^{(0)} \equiv 0$. In the first order approximation, by
(\ref{Phi_1_solution}), $G^{(1)}$ contains the derivatives of $\alpha_j^{(0)} \equiv \alpha_j, j =
1,2,$ up to the first order:
\begin{equation}
G^{(1)} \equiv -i \left( \frac{\partial \alpha_1}{\partial \xi} - \frac{\partial \alpha_2}{\partial
\eta}\right) \frac{\partial \bvph^H_*}{\partial p_x} -i \left( \frac{\partial \alpha_1}{\partial
\eta} + \frac{\partial \alpha_2}{\partial \xi}\right) \frac{\partial \bvph^E_*}{\partial p_x}-
i\frac{\partial \alpha_1}{\partial \zeta} \frac{\partial \bvph^X_*}{\partial p_z} - i\frac{\partial
\alpha_2}{\partial \zeta} \frac{\partial \bvph^Y_*}{\partial p_z}.
\end{equation} }
To find the approximation $\bphi^{(2)}$ we consider the following inhomogeneous equations:
\begin{eqnarray}
\A_* \Upsilon^{2H}_{j} = \G_j \frac{\partial \bvph^H_*}{\partial p_x},&& \quad \A_*
\Upsilon^{2E}_{j} = \G_j\frac{\partial
\bvph^E_*}{\partial p_x},\label{eq-upsilon1} \\
\A_* \Upsilon^{2X}_{j} = \G_j \frac{\partial \bvph^X_*}{\partial p_z},&& \quad \A_*
\Upsilon^{2Y}_{j} = \G_j\frac{\partial \bvph^Y_*}{\partial p_z}, \quad j=1,2,3.\label{eq-upsilon2}
\end{eqnarray}
Each of these equations is a system of the form (\ref{Maxwell-series}) with nonzero right-hand sides. It is
necessary to check the solvability of these equations in the class $\mathcal{M}$. To do this, we check the conditions imposed in  Lemma 2. This means that the conditions (\ref{lemma2}) must be satisfied:
\begin{equation}
\left(\bvph^X_*,\vec{F}\right)=0,\quad \left(\bvph^Y_*,\vec{F}\right)=0,
\end{equation}
where $\vec{F}$ stands for each expression on the right-hand sides of the equations
(\ref{eq-upsilon1}), (\ref{eq-upsilon2}). By taking all the possible combinations of the form
\begin{equation}
\begin{aligned}
& \left(\bvph^f_*, \G_j \frac{\partial \bvph^H_*}{\partial p_x} \right)=0, \quad \left(\bvph^f_*,
\G_j\frac{\partial
\bvph^E_*}{\partial p_x} \right)=0,  \\
&\left(\bvph^f_*, \G_j \frac{\partial \bvph^X_*}{\partial p_z} \right)=0, \quad \left(\bvph^f_*,
\G_j \frac{\partial \bvph^Y_*}{\partial p_z} \right)=0, \,\, j = 1,2,3, \quad f = X,Y,
\end{aligned}
\end{equation}
we obtain 24 conditions, 18 of them are satisfied (which follows from (\ref{help1}), (\ref{help2}),
(\ref{g3d})), and 6 are not satisfied (which follows from (\ref{om2-1}), (\ref{om2-2}) and
(\ref{om2-3})). Nonhomogeneous equations from the set (\ref{eq-upsilon1}), (\ref{eq-upsilon2}),
for which  the solvability conditions are satisfied, have  solutions in the class $\mathcal{M}$.
The other equations also have  solutions, which do not belong to the class $\mathcal{M}$ and are
nonperiodic.

Next we take the sum of the equations for $\Upsilon^{2H}_{j}$, $\Upsilon^{2E}_{j}$,
$\Upsilon^{2X}_{j}$, $\Upsilon^{2Y}_{j}$, $j=1,2,3,$ with  coefficients such that the sum of the
right-hand side expressions coincides with $F^{(2)} \equiv -i\hG \cdot\nabla_{\brho}\bphi^{(1)}$.
The solution of the obtained  sum of equations
%The terms in the
%right-hand side of the equations, which do not have solutions from the class $\mathcal{M}$, will
%form a linear combination with such coefficients, that it coincides with equations (\ref{eq0-1})
%and (\ref{eq0-2}) on $\tau_1, \tau_2$, which we assume to be satisfied. The non-periodicity of the
%solutions will be compensated, and
%The sum of the solutions of the equations (\ref{eq-upsilon1}),
%(\ref{eq-upsilon2})
  belongs to the class $\mathcal{M}$, because the conditions of solvability, which are reduced to (\ref{eq0-sys}),  are  satisfied. Thus, we get
\begin{equation}\label{Phi_2_G}
\begin{aligned}
G^{(2)} \equiv &-i \left( \frac{\partial \alpha_1^{(1)}}{\partial \xi} - \frac{\partial
\alpha_2^{(1)}}{\partial \eta}\right) \frac{\partial \bvph^H_*}{\partial p_x} -i \left(
\frac{\partial \alpha_1^{(1)}}{\partial \eta} + \frac{\partial \alpha_2^{(1)}}{\partial \xi}\right)
\frac{\partial
\bvph^E_*}{\partial p_x}- \\
& -i\frac{\partial \alpha_1^{(1)}}{\partial \zeta} \frac{\partial \bvph^X_*}{\partial p_z} -
i\frac{\partial \alpha_2^{(1)}}{\partial \zeta} \frac{\partial \bvph^Y_*}{\partial p_z}-
\\
& -\sum\limits_{j=1}^3\nabla_j \G_j \left[ \left( \frac{\partial \alpha_1}{\partial \xi} -
\frac{\partial \alpha_2}{\partial \eta}\right) \Upsilon^{2H}_{j}(z) + \left( \frac{\partial
\alpha_1}{\partial
\eta} + \frac{\partial \alpha_2}{\partial \xi}\right) \Upsilon^{2H}_{j}(z)\right.+ \\
& \left. + \frac{\partial \alpha_1}{\partial \zeta}\Upsilon^{2X}_{j} + \frac{\partial
\alpha_2}{\partial \zeta}\Upsilon^{2Y}_{j}\right],\quad \nabla_1 = \frac{\partial}{\partial \xi},
\nabla_2 = \frac{\partial}{\partial \eta}, \nabla_3 = \frac{\partial}{\partial \zeta}.
\end{aligned}
\end{equation}

 In an analogous manner, we can derive that the term $G^{(3)}$ contains the first derivatives of $\alpha_1^{(2)}$ and $\alpha_2^{(2)}$, the
second derivatives of $\alpha_1^{(1)}$ and $\alpha_2^{(1)}$, and the third derivatives of
$\alpha_1$ and $\alpha_2$. The approximation $\bphi^{(k)}$ contains yet unknown functions
$\alpha^{(k)}_j, \, j=1,2$, and $G^{(k)}$, which depends on the first derivatives
of $\alpha_1^{(k-1)}$ and $\alpha_2^{(k-1)}$, the second derivatives of $\alpha_1^{(k-2)}$ and
$\alpha_2^{(k-2)}$, and the derivatives of the $k$th order of $\alpha_1$ and $\alpha_2$. The
solvability conditions of $\bphi^{(k+1)}$,
\begin{equation}\label{solv-n}
\left(\bvph^X_*,\hG\cdot\nabla_{\brho} \bphi^{(k)}\right)=0,\quad
\left(\bvph^Y_*,\hG\cdot\nabla_{\brho} \bphi^{(k)}\right)=0,
\end{equation}
provide equations for $\alpha_1^{(k-1)}$ and $\alpha_2^{(k-1)}$. By analogy with the section
\ref{sec_second_approx}, we obtain a system of  partial differential equations:
\begin{eqnarray}
&& \frac{\partial^2 \alpha^{(k-1)}_1}{\partial \xi^2} \ddot{\omega}_{11*}^H + \frac{\partial^2
\alpha^{(k-1)}_1}{\partial \eta^2} \ddot{\omega}_{11*}^E + \frac{\partial^2
\alpha^{(k-1)}_1}{\partial \zeta^2}\ddot{\omega}_{33*}^0  + 2\frac{\delta
\omega}{c}\alpha^{(k-1)}_1-  \nonumber
\\ && - \frac{\partial^2 \alpha^{(k-1)}_2}{\partial \xi \partial \eta}
(\ddot{\omega}_{11{*}}^H-\ddot{\omega}_{11{*}}^E)= A_1^{(k-1)}\left(\alpha_1^{(k-2)},
\alpha_2^{(k-2)}, \ldots, \alpha_1, \alpha_2 \right),
\end{eqnarray}
\begin{eqnarray}
&& \frac{\partial^2 \alpha^{(k-1)}_2}{\partial \xi^2} \ddot{\omega}_{11*}^E + \frac{\partial^2
\alpha^{(k-1)}_2}{\partial \eta^2} \ddot{\omega}_{11*}^H  + \frac{\partial^2
\alpha^{(k-1)}_2}{\partial \zeta^2}\ddot{\omega}_{33*}^0  + 2\frac{\delta
\omega}{c}\alpha^{(k-1)}_2- \nonumber
\\&& - \frac{\partial^2 \alpha^{(k-1)}_1}{\partial \xi \partial
\eta}(\ddot{\omega}_{11{*}}^H-\ddot{\omega}_{11{*}}^E)=A_2^{(k-1)}\left(\alpha_1^{(k-2)},
\alpha_2^{(k-2)}, \ldots, \alpha_1, \alpha_2 \right), \label{al-higher}
\end{eqnarray}
where $A_1^{(0)}=A_2^{(0)}=0$, $A_1^{(k-1)}, A_2^{(k-1)}$ for $k>1$ are  linear combinations of
 derivatives of the already known functions $\alpha_j^{(l)}, j = 1,2, \, l =1, \ldots, k-2,$ with
respect to the variables $\xi,\eta,$ and $\zeta$. This means that equations for approximations of all
orders (\ref{Maxwell-series}-\ref{Fn}) can be solved step by step.

\section{Solution of the equations for $\alpha$}\label{sec:alpha}

We have obtained the equations (\ref{eq0-sys}) with constant coefficients, which describe the
behavior of the envelopes of the field. Now we are going to discuss the methods of solving them.
The simplest case arises if the field does not depend on one of the lateral coordinates, for
example, on $\eta$. In this case, the equations for $\alpha_1$ and $\alpha_2$ are separated:
\begin{equation}\label{twoD-al}
\begin{aligned}
&\frac{\partial^2 \alpha_1}{\partial\xi^2}\ddot{\omega}_{11*}^H  + \frac{\partial^2
\alpha_1}{\partial \zeta^2}\ddot{\omega}^0_{33*} + 2
\frac{\delta\omega}{c}\alpha_1 = 0,  \\
&\frac{\partial^2 \alpha_2}{\partial\xi^2}\ddot{\omega}_{11*}^E  + \frac{\partial^2
\alpha_2}{\partial \zeta^2}\ddot{\omega}^0_{33*} + 2 \frac{\delta\omega}{c}\alpha_2 = 0,
\end{aligned}
\end{equation}
where the coefficients are defined in (\ref{omega_dot_dot}) and are the derivatives of the
dispersion functions for TM and TE type waves, calculated at stationary points. Equations
(\ref{twoD-al}) can be elliptic or hyperbolic ones depending on the type of a stationary point. According to the Appendix 1, we have $\ddot{\omega}_{11*}^f>0$, $f=E,H$.
% {as it follows from (\ref{App1-omx}), $\ddot{\omega}_{11*}^f>0$, $f=E,H$}.
Equations (\ref{twoD-al}) are
elliptic if $\ddot{\omega}^0_{33*}
>0$, and hyperbolic if $\ddot{\omega}^0_{33*}<0$. We solve these equations by means of the Fourier
method performing the Fourier transform with respect to the variable $\xi$.  {By dividing the obtained equations by
the coefficient $\ddot{\omega}^0_{33*}$,} we derive the following equations for the functions
$\widehat{\alpha}_j(p_{\xi},\zeta):$
\begin{equation}\label{eq1111}
\frac{\partial^2 \widehat{\alpha}_j}{\partial \zeta^2} -
\frac{\ddot{\omega}^j_{11*}}{\ddot{\omega}^0_{33*}}p_{\xi}^2 \widehat{\alpha}_j + 2\frac{\delta
\omega}{c\,\ddot{\omega}^0_{33*}}\widehat{\alpha}_j  = 0, \quad  {j=1,2},
\end{equation}
\begin{equation}\label{def12HE}
\ddot{\omega}^1_{11*} \equiv \ddot{\omega}^H_{11*}, \quad \ddot{\omega}^2_{11*} \equiv \ddot{\omega}^E_{11*}.
\end{equation}
Solutions of (\ref{eq1111}) are determined by the integral
\begin{equation}
\alpha_j(\brho) = \frac{1}{2\pi} \int\limits_{\mathbb{R}} dp_{\xi} \,\, e^{ip_{\xi}\xi} \left[
\widehat{\alpha}^-_j(p_{\xi}) \, e^{ip_{j\zeta}\zeta} + \widehat{\alpha}^+_j(p_{\xi}) \,
e^{-ip_{j\zeta}\zeta} \right],
\end{equation}
where  $\brho = (\xi,0,\zeta)$ and
\begin{equation}
p_{j\zeta} = \sqrt{-p_{\xi}^2 \frac{\ddot{\omega}^j_{11*}}{\ddot{\omega}^0_{33*}} + 2\frac{\delta
\omega}{c\,\ddot{\omega}^0_{33*}}}.
\end{equation}
If $\ddot{\omega}^0_{33*} > 0$ and $0 < p_{\xi}^2\ddot{\omega}^j_{11*}/\ddot{\omega}^0_{33*} <
2\delta \omega/c\,\ddot{\omega}^0_{33*}$, then the integral describes the propagating waves
governed by the elliptic equation. The components with  such values of $p_{\xi}$ that
$p_{\xi}^2\ddot{\omega}^j_{11*} > {2}\delta \omega/c$ do not propagate. If $\ddot{\omega}^0_{33*}
< 0$, then  introducing the notation $\sigma^2_j$,  $q^2$, we obtain
\begin{equation}\label{sigma}
p_{j\zeta} = \sqrt{p_{\xi}^2\sigma^2_j + q^2}, \quad \sigma^2_j =
\frac{\ddot{\omega}_{11*}^j}{|\ddot{\omega}^0_{33*}|}, \quad q^2 = -2 \frac{\delta\omega}{c
|\ddot{\omega}^0_{33*}|}.
\end{equation}
  {If $\delta\omega>0$, only the components with large $p_{\xi}$ propagate.}
%In this case, only high-frequency components propagate, while low-frequency components stop.
The equation (\ref{eq1111}) turns to a Klein-Gordon-Fock type equation.

If $\delta\omega = 0$, the equations (\ref{twoD-al}) are the one-dimensional wave equations, where
the variable $\zeta$ plays the role of time, and $\sigma$ plays the role of the speed of wave
propagation. The solutions can be found by the D'Alambert method and read
\begin{equation}
\alpha_j = F_j(\xi - \sigma\zeta) + G_j(\xi + \sigma\zeta),
\end{equation}
where $F_j$ and $G_j$ are some functions. If the function $F_j$ is localized for  $\zeta=0$
near $\xi=0$, it is localized for any $\zeta$ near the line $\xi - \sigma\zeta = 0$  and
propagates undistorted. This means that in the medium under consideration there is the possibility
of existence of non-distorting beams, all of them having the the same angle with $z$ axis equal
to
\begin{equation}\label{angle_dif}
\varphi^f = \pm\mathrm{arctg} \sqrt{\frac{\ddot{\omega}_{11*}^f}{|\ddot{\omega}^0_{33*}|}}, \qquad
f=H, E.
\end{equation}

%Thus, we obtain un undistorted beam on the $(\xi,\zeta)$ plane.
We have discussed and studied this effect numerically in \cite{PerelSidorenko2012}.

Now we proceed to the general case, where the field depends on both lateral coordinates $\xi, \eta$.
We obtained equations (\ref{eq0-1}) and (\ref{eq0-2}) for the functions $\tau_1, \tau_2$ defined by
(\ref{tau-def}). Let us take the derivative of the equation (\ref{eq0-1}) with respect to $\xi$ and
of the equation (\ref{eq0-2}) with respect to $\eta$ and then calculate the difference. We obtain
an equation for $\tau_1$. To get the equation for $\tau_2$, we differentiate (\ref{eq0-1}) with
respect to $\eta$ and equation (\ref{eq0-2}) with respect to $\xi$ and take the sum of the results.
Thus, we find the equations for the functions $\tau_1, \tau_2$
\begin{equation}
\begin{aligned}\label{tau-3Deq}
\ddot{\omega}_{11*}^{H} \left( \frac{\partial^2 \tau_1}{\partial \xi^2} + \frac{\partial^2
\tau_1}{\partial \eta^2} \right) + \ddot{\omega}^0_{33*}\frac{\partial^2 \tau_1}{\partial \zeta^2}
+ 2\frac{\delta\omega}{c}\tau_1 = 0, \\
\ddot{\omega}_{11*}^{E} \left( \frac{\partial^2 \tau_2}{\partial
\xi^2} + \frac{\partial^2 \tau_2}{\partial \eta^2} \right) +
\ddot{\omega}^0_{33*}\frac{\partial^2 \tau_2}{\partial \zeta^2} +
2\frac{\delta\omega}{c}\tau_2 = 0.
\end{aligned}
\end{equation}
These equations are either hyperbolic or elliptic ones depending on the sign of
$\ddot{\omega}^0_{33*}$. Their solutions obtained by the Fourier transform read
\begin{equation}\label{tau_sol_four}
\tau_j(\brho) = \frac{1}{(2\pi)^2}\int\limits_{\mathbb{R}^2} dp_{\xi} dp_{\eta} \,
e^{ip_{\xi}\xi+ip_{\eta}\eta} \left[ \widehat{\tau}^{+}_j(p_{\xi}, p_{\eta}) e^{-ip_{j\zeta} \zeta}
+ \widehat{\tau}^{-}_j(p_{\xi}, p_{\eta}) e^{ip_{j\zeta} \zeta} \right],
\end{equation}
where
\begin{equation}
p_{1\zeta} = \sqrt{-\frac{\ddot{\omega}^H_{11*}}{\ddot{\omega}^0_{33*}}(p_{\xi}^2 + p_{\eta}^2) +
2\frac{\delta \omega}{c\,\ddot{\omega}^0_{33*}}}, \qquad p_{2\zeta} =
\sqrt{-\frac{\ddot{\omega}^E_{11*}}{\ddot{\omega}^0_{33*}}(p_{\xi}^2 + p_{\eta}^2) + 2\frac{\delta
\omega}{c\,\ddot{\omega}^0_{33*}}},
\end{equation}
 {and $\widehat{\tau}^{\pm}_j(p_{\xi}, p_{\eta})$ are such functions that the integrals
(\ref{tau_sol_four}) and the integrals, obtained by taking the derivative of (\ref{tau_sol_four}),
converge.} For $\ddot{\omega}^0_{33*} < 0$,  {$\delta\omega >0$ only the components with large
$p_{\xi}^2+p_{\eta}^2$ propagate.}

Now we are going to derive the equations for the original functions $\alpha_1, \alpha_2$. We take
the derivatives of two relations (\ref{tau-def}) with respect to $\eta$ and $\xi$, respectively,
and take their sum and difference. We obtain
\begin{equation}
\triangle \alpha_1 = \frac{\partial \tau_1}{\partial \xi} + \frac{\partial \tau_2}{\partial \eta},
\quad \triangle \alpha_2 = \frac{\partial \tau_2}{\partial \xi} - \frac{\partial \tau_1}{\partial
\eta}, \quad \triangle \equiv \frac{\partial^2 }{\partial \xi^2} + \frac{\partial^2 }{\partial
\eta^2}.
\end{equation}
By the Fourier integral we get the following expressions for the functions $\alpha_1, \alpha_2$:
\begin{equation}\label{alpha_1alpha_2}
\begin{aligned}
\alpha_1(\brho) = \frac{-i}{(2\pi)^2} \int\limits_{\mathbb{R}^2} dp_{\xi}dp_{\eta} \,
e^{i(p_{\xi} \xi + p_{\eta} \eta)}  \frac{p_{\xi}\widehat{\tau}_1 +
p_{\eta}\widehat{\tau}_2}{p_{\xi}^2 + p_{\eta}^2}, \\
\alpha_2(\brho) = \frac{-i}{(2\pi)^2} \int\limits_{\mathbb{R}^2} dp_{\xi}dp_{\eta} \,
e^{i(p_{\xi} \xi + p_{\eta} \eta)} \frac{p_{\xi}\widehat{\tau}_2 -
p_{\eta}\widehat{\tau}_1}{p_{\xi}^2 + p_{\eta}^2},
\end{aligned}
\end{equation}
where
\begin{equation}
\widehat{\tau}_j \equiv \widehat{\tau}^{+}_j(p_{\xi}, p_{\eta}) e^{-ip_{j\zeta} \zeta} +
\widehat{\tau}^{-}_j(p_{\xi}, p_{\eta}) e^{ip_{j\zeta} \zeta}.
\end{equation}
We also require the functions $\widehat{\tau}_j p_{\xi}/(p_{\xi}^2 + p_{\eta}^2)$ and
$\widehat{\tau}_j p_{\eta}/(p_{\xi}^2 + p_{\eta}^2)$ to be continuous  {and integrable}. The
expressions for $\alpha_1, \alpha_2$ can also be rewritten in the polar coordinate system
$(p_{\rho},\gamma)$ as
\begin{equation}\label{alpha_1alpha_2_polar}
\begin{aligned}
\alpha_1(\brho) = \frac{-i}{(2\pi)^2} \int\limits_0^{+\infty} dp_{\rho} \int\limits_{0}^{2\pi}
d\gamma \, e^{i(p_{\rho}\xi \cos\gamma + p_{\rho}\eta \sin\gamma) }  (\cos
\gamma \widehat{\tau}_1 + \sin\gamma \widehat{\tau}_2), \\
\alpha_2(\brho) = \frac{-i}{(2\pi)^2} \int\limits_0^{+\infty} dp_{\rho} \int\limits_{0}^{2\pi}
d\gamma \, e^{i(p_{\rho}\xi \cos\gamma + p_{\rho}\eta \sin\gamma) } (\cos\gamma \widehat{\tau}_2 -
\sin\gamma \widehat{\tau}_1),
\end{aligned}
\end{equation}
 {where $p_{\rho} = \sqrt{p_{\xi}^2 + p_{\eta}^2}$, $\cos{\gamma} = p_{\xi}/p_{\rho}$,
 $\sin{\gamma} = p_{\eta}/p_{\rho}$}.

We express also the principal approximation of the asymptotic solutions in terms of TM and TE solutions at the stationary point determined by the passage to the limit in every direction (\ref{lim-old}).
Substituting solution of (\ref{eq0-sys}) given by     (\ref{alpha_1alpha_2_polar}) in formula  (\ref{princip}), we obtain the integral representation
\begin{equation}
\left( \begin{array}{c} {\mathbf E} \\ {\mathbf{H}} \end{array}
\right)(z, \brho)  \simeq \frac{-i}{(2\pi)^2} \int\limits_0^{+\infty} dp_{\rho}
\int\limits_{0}^{2\pi} d\gamma \, e^{i(p_{\rho}\xi \cos\gamma + p_{\rho}\eta \sin\gamma) }  ( \widehat{\tau}_1  \bPh^{H}_*(z)  +  \widehat{\tau}_2 \bPh^{E}_*(z) ),
\end{equation}
where $\bPh^{H}_*(z)$ and $\bPh^{E}_*(z)$ depend on the direction of propagation  according to (\ref{lim-old}).

\section{Conclusions}

We have elaborated a formal asymptotic approach for monochromatic electromagnetic fields in a layered periodic structure. The frequency of the field is close to that of a stationary point $\vp_*$  of one of the sheets of the dispersive surface $\omega=\omega^f(\vp)$. Here $f$ stands for the type of the polarization, which may be $H$
for TM- or $E$ for TE- polarization. The dispersive surfaces for waves of different polarizations are distinct, but the stationary points of them coincide.
For the conditions listed in Section \ref{two-scale}, we found  asymptotic series for the solutions of Maxwell equations (\ref{Maxwell}) by the two-scale expansions method.
 The field is assumed to be a function of a fast variable $z$ and slow variables $\rho$; see (\ref{Psi-form}).
  The field in the principal approximation  is represented as a linear combination of  Floquet-Bloch solutions  of different
polarizations  $\bPh_*^f(z),$ $f=X$ or $Y$  with  slowly varying envelopes $\alpha_{j}(\brho),$ $j=1,2$. It reads
\begin{equation}\label{princip-final}
\bPs = \left(
\begin{array}{c} \mathbb{E} \\ \mathbb{H} \end{array} \right)(z, \brho) \simeq \alpha_{1}(\brho)\bPh^{X}_*(z) +
\alpha_{2}(\brho) \bPh^{Y}_*(z),
\end{equation}
where
\begin{equation}
 \bPh_*^f(z) \equiv \bPh^f(z,\vp_*) =  e^{ip_{z*}z}\bvph_*^f, \quad  f=X,Y,
\end{equation}
where $\vp_* =(0,0, p_{z*})$, $p_{z*} = 0, \pm\pi/b.$  The functions $\bPh^f(z,\vp)$
  are  Floquet-Bloch solutions
of the system (\ref{Maxwell-pxpy}). For $\vp=\vp_*$,  they are expressed  in terms of $(E_0, H_0)$ (see, (\ref{Bas-XY})), which are  Floquet-Bloch solutions of the system (\ref{E0-H0}) with the quasimomentum $p_{z*}$.

The envelope functions $\alpha_j, \,\, j=1,2$, are defined by the
 equations with constant coefficients (\ref{eq0-sys}).
These coefficients are the second derivatives of the dispersion functions, i.e., the coefficients of the Tailor
expansion of $\omega=\omega^f(\vp)$ near the stationary points
$\vp_*=(0,0,p_{z*})$
\begin{equation}
\omega=\omega_*+ \frac{1}{2}\ddot{\omega}_{11*}^{f} (p_x^2+p_y^2) +
\frac{1}{2}\ddot{\omega}^0_{33*} (p_z-p_{z*})^2 + \ldots,
\end{equation}
where $\ddot{\omega}_{11*}^{f}$, $\ddot{\omega}^0_{33*}$ are the
second derivatives of the dispersion function $\omega$ with respect
to $p_x$ and $p_z$ calculated at the stationary point. Since the
problem is axially symmetric, the derivatives of the dispersion
functions with respect to $p_x$ and $p_y$ are equal and depend on
the type of the polarization, while the derivatives with respect to
$p_z$ do not depend on the type of the polarization.

The system (\ref{eq0-sys}) can be split by introducing new functions $\tau_j$, $j=1,2,$ by means of (\ref{tau-def}).
As  was obtained in Section \ref{sec:alpha}, the functions $\tau_1$ and $\tau_2$ satisfy the equations (\ref{tau-3Deq}).
The coefficients $\alpha_j, j=1,2,$ are expressed in terms of both  $\tau_j, j=1,2,$ in (\ref{alpha_1alpha_2_polar}). The system can be split into two separated equations only if the field does not depend on one of the spatial coordinates. Such analysis was absent
in the  papers of Longhi \cite{Longhi}, \cite{LonghiJanner}.
%If $\tau_j, \,\, j=1,2$, the functions $\alpha_j\,\, j=1,2$, are found via the equations
%\begin{equation}
%\triangle \alpha_1 = \frac{\partial \tau_1}{\partial \xi} +
%\frac{\partial \tau_2}{\partial \eta}, \quad \triangle \alpha_2 =
%\frac{\partial \tau_2}{\partial \xi} - \frac{\partial
%\tau_1}{\partial \eta}, \quad \triangle \equiv \frac{\partial^2
%}{\partial \xi^2} + \frac{\partial^2 }{\partial \eta^2}.
%%\end{equation}

 An interesting particular case arises if $\ddot{\omega}^0_{33*} < 0$. Then the equations for $\tau_{1}$ and
$\tau_2$ are  hyperbolic of the Klein-Gordon-Fock type. The coordinate
$\zeta$ stands for  time. If additionally  $\delta\omega = 0$, there are wave equations. The
effects that arise if the field depends only on one lateral coordinate were studied both
analytically and numerically in our paper \cite{PerelSidorenko2012}.

 The  investigation of  qualitative consequences of the obtained results in the case of two lateral coordinates are out of  scope of the present paper and will be discussed in the next publications. We expect that,
 by choosing the localized solutions of envelopes, we can construct
beam-like solutions of the Maxwell equations.
The obtained formulas  enable us also to find a change of  polarization of the field in passing through the layered periodic structure.

The results may be generalized to another equations, which can be written in  matrix form (\ref{Maxwell}).

\section*{Appendix 1}

\subsection*{The dispersion relation} To make the paper self-contained, we obtain the dispersion
relation and find  stationary points of the dispersive surfaces. To do this, we consider the Floquet-Bloch solutions of the
first subsystems (\ref{Maxwell-TM}):
\begin{equation}\label{EH-Bl}
\left(\begin{array}{c} E_{\parallel}\\
H_{\perp} \end{array}\right) = e^{ipz} \left(\begin{array}{c} U_1\\
U_2 \end{array}\right),
\end{equation}
 where $U_1, U_2$ are the periodic functions of $z$ with period $b$.

We accomplish the  {first of subsystems} (\ref{Maxwell-TM}) with the initial data
$E_{\parallel}=1, H_{\perp}=0$ and denote the solution of such a Cauchy problem by $e_1,h_1$. We
introduce another solution $e_2,h_2$, which satisfies (\ref{Maxwell-TM}) and the initial data
$E_{\parallel}=0, H_{\perp}=1$. Both solutions are smooth in the intervals, where $\varepsilon,\mu$
are continuous, and continuous at the points of discontinuity of the parameters $\varepsilon,\mu$, but,
generally speaking, they are not periodic.  These solutions depend on $p_{\parallel}^2, \omega$,
because the coefficients of (\ref{Maxwell-TM}) depend on these parameters. These solutions are
linear independent and form a basis in the space of the solutions of the first subsystem of
(\ref{Maxwell-TM}). We introduce the matrix $\mathbf{M}$ of the solutions $(e_1,h_1)^t$ and
$(e_2,h_2)^t$ as follows:
\begin{equation}
\mathbf{M}(z;  p_{\parallel}^2, \omega) = \left(\begin{array}{cc} e_1 & e_2\\
h_1 & h_2 \end{array}\right)(z;  p_{\parallel}^2, \omega).
\end{equation}
The matrix $\mathbf{M}(b; p_{\parallel}^2, \omega)$ is then a monodromy matrix:
\begin{equation}
\mathbf{M}(z+b;  p_{\parallel}^2, \omega)=\mathbf{M}(z;  p_{\parallel}^2, \omega)\mathbf{M}(b;
p_{\parallel}^2, \omega).
\end{equation}
%and the solution $(E_{\parallel}, H_{\perp})^t$ in this notation reads
 {We seek the Floquet-Bloch solutions in the following form:}
\begin{equation}
\left(\begin{array}{c} E_{\parallel}\\
H_{\perp} \end{array}\right) = \mathbf{M}(z;  p_{\parallel}^2, \omega) \left(\begin{array}{c} \beta_1\\
\beta_2 \end{array}\right),\label{App-Floq}
\end{equation}
 where $(\beta_1, \beta_2)^t$ is the eigenvector corresponding to the eigenvalue $\lambda$ of the problem:
\begin{equation}\label{eigenvec}
\left(\mathbf{M}(b;  p_{\parallel}^2, \omega) - \lambda \mathbf{I} \right)\left(\begin{array}{c}
\beta_1\\ \beta_2
\end{array}\right) = \left(\begin{array}{c} 0\\ 0
\end{array}\right),
\end{equation}
here $\mathbf{I}$ is the identity matrix.  {The equation for $\lambda$ is expressed in terms of
the determinant and the trace of the matrix $\mathbf{M}$}. The determinant of the matrix
$\mathbf{M}(b;  p_{\parallel}^2, \omega)$ is the Wronskian of the solutions $(e_1,h_1)^t$ and $(e_2,h_2)^t$ of the system (\ref{Maxwell-TM}) at $z=b$. It is
constant by the Ostrogradskiy-Liuville theorem and, thus, can be calculated for any $z$. For $z=0$
it is equal to 1, so $\mathrm{det}\, \mathbf{M}(b;  p_{\parallel}^2, \omega) = 1$ and $\lambda$
satisfies the quadratic equation
\begin{equation}\label{quadric}
\lambda^2 + {\rm Sp} \mathbf{M}(b;  p_{\parallel}^2, \omega) \lambda + 1 = 0.
\end{equation}
In order to obtain the Floquet-Bloch solutions, we require  $|\lambda|=1$ and we may assume
$\lambda_1 = e^{i p_z b}$, where $p_z$ is real-valued. Then $\lambda_2 = e^{-i p_z b}$ and
\begin{equation}
\label{disp1} M_{11}(b;  p_{\parallel}^2, \omega) + M_{22}(b;  p_{\parallel}^2, \omega) = 2
\cos{(p_z b)}.
\end{equation}
 This equation establishes the relation between $p_z$, $p_{\parallel}^2$ and $\omega$, so we consider below only
two of these three variables as  free parameters. We denote $\mathcal{F} \equiv M_{11}(b;
p_{\parallel}^2, \omega) + M_{22}(b;  p_{\parallel}^2, \omega)$. The function $\mathcal{F} \equiv
{\cal F}( p_{\parallel}^2, \omega)$ depends on the problem parameters $\omega, p_{\parallel}$
analytically, because the coefficients of the system (\ref{Maxwell-TM}) depend on these parameters
analytically if $\omega \neq 0$. Therefore the dispersion relation (\ref{disp1}) can be written in
the form of
\begin{equation}\label{disp2}
{\cal F}( p_{\parallel}^2, \omega) -  \cos{(p_z b)}=0.
\end{equation}
The function ${\cal F}$ is an oscillating real-valued function of $\omega$, its minima and maxima
are larger than or equal  $1$ (see, for example, \cite{Eastham}).  The corresponding $p_z$
changes from $-\pi/b$ to $\pi/b$. If $|\mathcal{F}|$ is greater than $1$ for some interval of
$\omega$, then the bounded solutions do not exist and such interval is called the forbidden zone.
For each interval of $\omega$ (the allowed zone), where $|\mathcal{F}|\leq 1$ and the function $\mathcal{F}$ of $\omega$  is monotone, there exists an inverse
function
\begin{equation}\label{omegaH}
\omega=\omega_j^H(\vp).
\end{equation}
If $|\mathcal{F}|=1$ and $\partial \mathcal{F}/\partial \omega=0$, two intervals of monotone
behavior of $\mathcal{F}$ touch each other at this point.  {Further, we assume that $\partial
\mathcal{F}/\partial \omega \ne 0.$ This follows from the assumption that the system has one
bounded and one unbounded solutions on the boundary of the interval of the monotone behavior of $\cal{F}$;
see, for example, \cite{Eastham}.}

Analogous considerations for the second subsystem of (\ref{Maxwell-TM}) yield the dispersion relation
$\omega=\omega_j^E(\vp)$ for the wave of the TE polarization.

In other words, the problem under consideration can be treated as a
spectral problem for the operator
\begin{equation}
\left(-i\textbf{P}^{-1}\G \cdot \nabla + \textbf{P}^{-1}(\vp \cdot \G)\right) \bvph =
\frac{\omega}{c_0} \bvph, \quad \bvph\in\mathcal{M},
\end{equation}
where $\omega$ plays the role of the spectral parameter and $\vp$ is an external parameter of the
problem. For fixed $\vp$, the operator has a discrete set of eigenvalues $\omega=\omega_j$. For $\vp\in
\mathbb{R}^2\times (-\pi/b, \pi/b)$, $\omega_j(\vp)$ forms a sheet of number $j$.

\subsection*{The stationary points of the dispersive surface}
We are going to find at least one stationary point of a single-valued function $\omega_j^H$,  where
$j$ is the number of the sheet of the dispersive surface. Below we omit $j$ for the sake of brevity.  We consider only such
zones, where $\partial \mathcal{F}/\partial \omega \neq 0$ on the entire zone. We take the
derivative of the implicitly defined function
\begin{equation}\label{deriv-omega}
\frac{\partial \omega^H}{\partial p_{\parallel}} = -2p_{\parallel}\left. \frac{\partial
\mathcal{F}}{\partial (p_{\parallel}^2)}\right/ \frac{\partial \mathcal{F}}{\partial \omega},
\qquad \frac{\partial \omega^H}{\partial p_z} = - b\sin p_zb\left/ \frac{\partial
\mathcal{F}}{\partial \omega}\right..
\end{equation}
We note that $\partial \omega^H/\partial p_z$ vanishes for $p_z
=0, \pm\pi/b$. The function $\omega^H(\vp)$ is a periodic function of $p_z$
with period $2\pi/b$. The derivative $\partial \omega^H/\partial
p_{\parallel}$ vanishes at least at the point $p_{\parallel}=0$
since $\mathcal{F}$ depends on $p_{\parallel}^2$. We denote the
stationary points with asterisk subscript: $\omega_* = \omega^H(\vp_*)$,
$p_{x*}=p_{y*}=0$, $p_{z*} =0, \pm\pi/b$. We note that such stationary points are the
bounds of the forbidden zones since $\cos p_{z*}b =\pm 1$, and hence
$\mathcal{F} = \pm1$.

Now we consider the second derivatives of $\omega$. First, we find the second derivative with
respect to $p_z$ from the formula (\ref{deriv-omega}). At a stationary point it reads
\begin{equation}
\frac{\partial^2 \omega^H}{\partial p_z^2} = -b^2\cos p_{z*}b \left/ \frac{\partial
\mathcal{F}}{\partial \omega} \right. - b\sin p_{z*}b \frac{\partial }{\partial p_z} \, \left(
\frac{\partial \mathcal{F}}{\partial \omega} \right).
\end{equation}
The second term is equal to zero owing to (\ref{disp2}), and the first term has a constant nonzero
numerator and a denominator with the sign changing on each sheet of the function $\omega$. Now we
consider the second derivative of $\omega$ with respect to $p_x$ (or $p_y$). By (\ref{om2-1}), we get
\begin{equation}\label{App1-omx}
\frac{\ddot{\omega}^H_{11*}}{c}  u_*^{XX} = 2\left(\bvph_*^X,\G_1 \frac{\partial
\bvph^H_*}{\partial p_x}\right),
\end{equation}
 {where
\begin{equation} \left.\frac{\partial^2 \omega^H}{\partial p_x^2}\right|_{\vp_*} \equiv
\ddot{\omega}^H_{11*}.
\end{equation}}
By the formula (\ref{der-p1}) and the definition of the matrix (\ref{P-G1-G2}), we conclude that
the scalar product on the right-hand side of the formula (\ref{App1-omx}) is equal to the integral of
$|H_0|^2/(k\varepsilon)$ with respect to $z$ and hence it is positive if $\varepsilon>0$. Since
$u^{XX} >0$ for $\varepsilon>0,\mu>0$ , the second derivative $\ddot{\omega}^H_{11*} > 0$. This
means that the stationary points of $\omega$ are of two different types: hyperbolic and elliptic,
depending on the sign of the second derivative with respect to $p_z$.

\subsection*{The Floquet-Bloch amplitudes}

 We seek the Floquet-Bloch solutions in the form (\ref{App-Floq}), where $(\beta_1, \beta_2)^t$
is the eigenvector of the monodromy matrix (\ref{eigenvec}), which corresponds to the eigenvalue
$\lambda$. Up to an arbitrary constant factor, $\beta_1 = M_{12}, \beta_2 = M_{11} - \lambda_1$. We
denote the variable $U$ corresponding to the choice of the sign $\pm$ in the exponent $e^{\pm i p_z b}$ with a subscript $\pm$. We
consider TM-type solution and denote it by the superscript $H$. Thus,
\begin{equation}\label{sol}
\mathbf{U}_+^H = e^{-ip_{z}z} \left( M_{12}
\left(\begin{array}{c} e_1\\
h_1 \end{array}\right)  +  (M_{11} - \lambda_1)
\left(\begin{array}{c} e_2\\
h_2 \end{array}\right)\right).
\end{equation}
Here $(e_j,h_j)^t, j=1,2$ depends on the variable $z$ and on the parameters $p_{\parallel}^2,
\omega$. The monodromy matrix also depends on $p_{\parallel}^2, \omega$.  We assume that the Floquet-Bloch solutions are expressed in terms of $\vp$.
We note that $p_{\parallel}^2 = p_x^2 + p_y^2$ and take into account the dispersion relation (\ref{omegaH}). The formula (\ref{EH-Bl}) takes a form:
\begin{equation}\label{Floq-Bl-App}
\left(\begin{array}{c} E_{\parallel}\\
H_{\perp} \end{array}\right)(z;\vp) = e^{ip_zz}
\mathbf{U}_+^H(z;p_z,p_{\parallel}^2,\omega^H(\vp)), \quad
\mathbf{U}_+^H = \left(\begin{array}{c} U_1\\
U_2 \end{array}\right).
\end{equation}
The second Floquet-Bloch solution is obtained by replacing $p_z$ by $-p_z$, and $\mathbf{U}_+^H$ by  $\mathbf{U}_-^H$.

An exception is the case $p_z = p_{z*}$, $p_{z*}=0, \pm\pi/b.$  Each point $p_{z*}$ corresponds to the boundary of the forbidden zone, where $|\mathcal{F}(p_{\parallel}^2,\omega)|=1$.  Then
the double root of the dispersion equation (\ref{quadric}) is $\lambda_1 = \lambda_2 = \pm 1$. We assume that $M_{12}\neq 0$. Formulas (\ref{Floq-Bl-App}), (\ref{sol}) determine the unique, up to the constant factor, solution for $\vp=\vp_*$  (i.e., $p_{\parallel}=p_{\parallel*}=0$, $p_z = p_{z*}$).
We seek  the second linearly independent solution in the form of
\begin{equation}\label{app-2}
\left(\begin{array}{c} E_{\parallel 2}\\
H_{\perp 2} \end{array}\right)(z; \vp_*) =
\mathbf{M}(z; 0, \omega_*) \left(\begin{array}{c} \gamma_1\\
\gamma_2 \end{array}\right),
\end{equation}
where the coefficients are found from the equation
\begin{equation}\label{adjoint}
\left( \mathbf{M}(b; 0, \omega_*) - \lambda \mathbf{I} \right)\left(\begin{array}{c} \gamma_1\\
\gamma_2
\end{array}\right) = \left(\begin{array}{c} \beta_1\\ \beta_2
\end{array}\right).
\end{equation}
 Then the solution (\ref{app-2}) satisfies the relation
\begin{equation}
\left(\begin{array}{c} E_{\parallel 2}\\ H_{\perp 2}
\end{array}\right)(z+b; \vp_*) = \lambda\left(\begin{array}{c}
E_{\parallel 2}\\ H_{\perp 2} \end{array}\right)(z; \vp_*) +
\left(\begin{array}{c} E_{\parallel}\\
H_{\perp} \end{array}\right)(z; \vp_*),
\end{equation}
 and can be written as
\begin{equation}
\left(\begin{array}{c} E_{\parallel 2}(z; \vp_*)\\ H_{\perp 2}(z; \vp_*)
\end{array}\right) = e^{ip_{z*}z} \left[\frac{z}{\lambda b}
\mathbf{U}^H_+(z; p_{z*},0,\omega_*) + \mathbf{Q}^H(z; p_{z*}, \omega_*) \right],
\end{equation}
where $\mathbf{U}^H_+$, $\mathbf{Q}^H$ are periodic functions of the variable $z$, the function $\mathbf{U}^H_+$ is defined
by (\ref{sol}). We note that $\mathbf{U}^H_+$ and $\mathbf{U}^H_-$ are proportional  at the point $\vp_*$.

For  waves of the TE polarization,  the Floquet-Bloch amplitudes are obtained analogously.

\section*{Appendix 2}

Let $\mathcal{M}$ be a class of six-component vector-valued functions of $z$, which are periodic with  period $b$,
  piecewise  smooth on the period, and their components with numbers  $1,2,4,5$ are continuous.

({\small{A piecewise function is a function that can be broken into a finite number of distinct pieces and on each piece both the function and its derivative are continuous, even though the whole function may have a jump discontinuity  at  points between the pieces.}
})

\textbf{Lemma 1}.

 The operator $\mathcal{A}_*$ defined by (\ref{Bl-amp2}) and (\ref{A-star}) on the
class of functions $\mathcal{M}$ is symmetric, i.e., for any $\w, \vv \in\mathcal{M}$ the following
relation is valid:
\begin{equation}
\left(\vv, \mathcal{A}_* \w \right) = \left(\mathcal{A}_* \vv, \w \right).
\end{equation}
This fact follows from the definition of the operator $\mathcal{A}_*$. Since $\mathbf{P}$
is a real-valued matrix and $\vv,\w \in \mathcal{M}$, we derive by integration by parts that
\begin{equation}\label{app1-1}
\left(\vv, \mathcal{A}_* \w \right) = \left(\vv, k_*\mathbf{P}\w + i\G_3\frac{\partial \w}{\partial z} - p_{z*} \G_3 \w
\right) = \left(\mathcal{A}_*\vv, \w \right) +i \sum\limits_{m=0}^M [\,<\vv,\G_3 \w>\,]_{m},
\end{equation}
where $<.,.>$ is a scalar product dependent on $z$; see (\ref{scal-angle}). By the definition of $\G_3$, see (\ref{P-G1-G2}),  we get
 \begin{equation}\label{app-jump}
 <\vv(z),\G_3 \w(z)> = \overline{v_1(z)}w_5(z) - \overline{v_2}w_4(z) - \overline{v_4(z)} w_2(z) + \overline{v_5(z)}w_1(z).
 \end{equation}
 We denote   a jump of any scalar  function $h(z)$ at the point $z_m$ by
\begin{equation}\label{jump}
 [\, h\,]_{m} = h(z_m+0) -  h(z_m-0),\quad m \ne 0; \qquad [\, h\,]_0 = h(b) -  h(0).
\end{equation}
The summation over all discontinuities of $\vv$ and $\w$ is implied in (\ref{app1-1}).
 Since $\vv, \w \in\mathcal{M}$, the function $<\vv(z),\G_3 \w(z)>$ is continuous, the terms outside the integral in (\ref{app1-1})  vanish, and the
symmetry of $\mathcal{A}_*$ is proved.

Now we proceed to the proof of \textbf{Lemma 2}, which states that a solution from the class
$\mathcal{M}$ of the equation
\begin{equation}\label{problem-A}
\A_* \bphi =\vec{F}
\end{equation}
 exists if and only if the following conditions are satisfied:
\begin{equation}
\left(\bvph^X_*,\vec{F}\right)=0,\quad \left(\bvph^Y_*,\vec{F}\right)=0,
\end{equation}
where the operator $\A_*$ is defined by the formula (\ref{def-A*}), the functions
$\bvph^{X}_*(z),$ $\bvph^{Y}_*(z)$ are defined by (\ref{vph-XY-def}), (\ref{Bl-sol-amp}), and
(\ref{Bas-XY}) and belong to ${\cal M}$.

In order to prove the lemma, we take the scalar product of (\ref{problem-A}) and $\bvph_*^X$:
\begin{equation}\label{lemma-proof1}
\left( \bvph_*^X, \mathcal{A}_*\bphi \right) = \left(\bvph^X_*,\vec{F}\right).
\end{equation}
First, if a solution of (\ref{problem-A}) $\bphi \in \mathcal{M}$, then the scalar product, by the
symmetry of $\mathcal{A}_*,$ can be rewritten as $\left( \bvph_*^X, \mathcal{A}_*\bphi \right) =
\left( \mathcal{A}_*\bvph_*^X, \bphi \right)$, and since $\mathcal{A}_*\bvph_*^X = 0$, it follows that the
scalar product on the right-hand side of the equation (\ref{lemma-proof1}) is also equal to zero.

Now let $\left(\bvph^X_*,\vec{F}\right)=0$. The equation (\ref{problem-A})  for the
vector-valued function $\bphi=(\phi_1, \phi_2, \phi_3,$ $\phi_4, \phi_5, \phi_6)^T$ has the form
\begin{equation}
\begin{array}{ll} \left\{ \begin{array}{rcl} \displaystyle{ k\varepsilon \phi_1 +
i\frac{\partial \phi_5}{\partial z} -p_{z*}\phi_5 = F_1}  \\ \\ \displaystyle{ k\mu \phi_5 + i\frac{\partial \phi_1}{\partial z}  -p_{z*}\phi_1
= F_5} \\ \\ \displaystyle{ k\varepsilon \phi_3 = F_3}
\end{array} \right.  & \qquad \left\{ \begin{array}{rcl}  \displaystyle{ k\mu \phi_4 - i\frac{\partial \phi_2}{\partial z}  + p_{z*}\phi_2 = F_4} \\ \\  \displaystyle{ k\varepsilon \phi_2 -
i\frac{\partial \phi_4}{\partial z}  +p_{z*}\phi_4 = F_2} \\ \\ \displaystyle{ k\mu \phi_6 = F_6.}
\end{array} \right. \end{array}
\end{equation}
It splits into the pair of nonhomogeneous systems by the components with numbers 1,5 and 2,4,
respectively. The components with numbers 3 and 6 are found explicitly. We consider one of the
subsystems for the components 1,5, and the second system is considered similarly. The solution of
the subsystem can always be represented  as a sum
\begin{equation}\nonumber%\label{sol_ful}
\left(\begin{array}{c} \phi_1\\ \phi_5 \end{array}\right) = A \mathbf{U}^H_+(z; p_{z*},0,\omega_*) +
B \left[ \frac{z}{b\lambda}\mathbf{U}^H_+(z; p_{z*},0,\omega_*) + \mathbf{Q}^H(z; p_{z*}, \omega_*)\right] +
\left(\begin{array}{c} \widetilde{\phi}_1\\ \widetilde{\phi}_5 \end{array}\right),
\end{equation}
where  the first two terms are the Floquet-Bloch amplitudes satisfying
the homogeneous system,  $A,B$ are some coefficients dependent on slow variables as on
parameters, and $( \widetilde{\phi}_1, \widetilde{\phi}_5 )^t$
indicates the particular vector solution of the nonhomogeneous equation,
which in general does not belong to the class $\mathcal{M}$. We show now that this
solution  can be chosen continuous at all the points inside $(0,b)$. To get the continuity
 inside $(0,b)$, we note that the solution $( \widetilde{\phi}_1, \widetilde{\phi}_5 )^t$ is always smooth on the intervals, where the parameters
$\varepsilon,\mu$ are continuous. It can have jumps at the boundary points of these intervals. At boundary points inside the period $(0,b)$, the jumps can be compensated by addition of the solutions of homogeneous equation with appropriate coefficients, which  depend on   intervals. However,
at the end of the period, the solution $( \widetilde{\phi}_1, \widetilde{\phi}_5 )^t$ may differ from the value at the
beginning of the period with jump $([\widetilde{\phi}_1]_0, [\widetilde{\phi}_5]_0)^t$; see (\ref{jump}) for  explanation of the notation. Integrating by parts, we obtain
 the following expression:
\begin{eqnarray}\label{app2-2}
&&\left(\bvph^X_*\, \mathcal{A}_*\bphi \right) = \left(\mathcal{A}_*\bvph^X_*, \bphi \right) +
\overline{E_0(0;\vp_*)}\,\,[\bphi_5]_0 + \overline{H_0(0;\vp_*)}\,\,[\bphi_1]_0 \nonumber \\ &&= \overline{E_0(0;\vp_*)}\,\,[\bphi_5]_0 + \overline{H_0(0;\vp_*)}\,\,[\bphi_1]_0 =
\left(\bvph^X_*,\vec{F}\right)=0,
\end{eqnarray}
where
\begin{equation}
\left(\begin{array}{c}{[\phi_1]_0}\\ {[\phi_5]_0} \end{array}\right) = \frac{1}{\lambda}B\,\,\mathbf{U}^H_+(b; p_{z*},0,\omega_*) +  \left(\begin{array}{c} {[\widetilde{\phi}_1]_0}\\ {[\widetilde{\phi}_5]_0} \end{array}\right).
\end{equation}
The jump of one of the components $\widetilde{\phi}_1$ or $\widetilde{\phi}_5$ of the particular
solution  at the end of the period can always be compensated by an appropriate choice of the coefficient
$B$ of the nonperiodic solution of the homogeneous equation. The  jump of the other component  vanishes  by
(\ref{app2-2}) if $H_0(0;\vp_*)\neq 0$, $E_0(0;\vp_*) \neq 0$. Then the constructed solution to the problem
$\mathcal{A}_*\bphi=\vec{F}$ belongs to the class $\mathcal{M}$ and the lemma is proved. If $E_0(0;\vp_*)$ or
$H_0(0;\vp_*)$ is equal to zero, we take the beginning of the period at another point.

\section*{Acknowledgements}
This work was supported by RFBR grant 140200624 (M.V.P. and M.S.S.) and by SPbGU grant 11.38.263.2014 (M.V.P).

\bibliography{PerelSidorenko_2016}

\end{document}